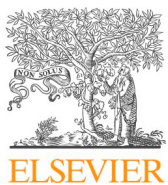
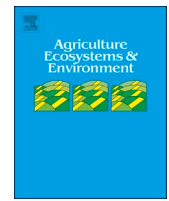

# Reducing greenhouse gas emissions and grain arsenic and lead levels without compromising yield in organically produced rice

Syed Faiz-ul Islam[a,b,d,*], Andreas de Neergaard[c], Bjoern Ole Sander[d], Lars Stoumann Jensen[b], Reiner Wassmann[d,e], Jan Willem van Groenigen[a]

[a] *Soil Biology Group, Wageningen University, Droevendaalsesteeg 3, PO Box 47, 6700 AA, Wageningen, the Netherlands*
[b] *Department of Plant and Environmental Sciences, University of Copenhagen, Thorvaldsensvej 40, DK-1871, Frederiksberg C, Denmark*
[c] *Faculty of Social Sciences, University of Copenhagen, Øster Farimagsgade 5, Copenhagen K, Denmark*
[d] *International Rice Research Institute (IRRI), Los Baños, Philippines*
[e] *Institute for Meteorology and Climate Research, Karlsruhe Institute of Technology (KIT), Garmisch-Partenkirchen, Germany*

ABSTRACT

Flooded rice production is crucial to global food security, but there are associated environmental concerns. In particular, it is a significant source of methane ($CH_4$) and nitrous oxide ($N_2O$) emissions and a large consumer of water resources, while arsenic levels in the grain are a serious health concern. There is also a tendency to use more organic fertilisers to close nutrient cycles, posing a threat of even higher GHG emissions and grain arsenic levels. It has been shown that alternate wetting and drying (AWD) water management reduces both water use and GHG emissions, but success at maintaining yields varies. This study tested the effect of early AWD (e-AWD) versus continuous flooding (CF) water management practices on grain yields, GHG emissions and grain arsenic levels in a split-plot field experiment with organic fertilisers under organic management. The treatments included: i) farmyard manure, ii) compost, and iii) biogas digestate, alone or in combination with mineral fertiliser. The e-AWD water regime showed no difference in yield for the organic treatments. Yields significantly increased by 5–16 % in the combination treatments. Root biomass and length increased in the e-AWD treatments up to 72 and 41 %, respectively. The e-AWD water regime reduced seasonal $CH_4$ emissions by 71–85 % for organic treatments and by 51–76 % for combination treatments; this was linked to a 15–47 % reduction in dissolved organic carbon (DOC), thereby reducing methanogenesis. $N_2O$ emissions increased by 23–305 % but accounted for <20 % of global warming potential (GWP). Area and yield-scaled GWPs were reduced by 67–83 %. The e–AWD regime altered soil redox potentials, resulting in a reduction in grain arsenic and lead concentrations of up to 66 % and 73 % respectively. Grain cadmium levels were also reduced up to 33 % in organic treatments. Structural equation modelling showed that DOC, redox, ammonium and root biomass were the key traits that regulated emissions and maintained yield. Despite the fact that the experiment was conducted in the dry-season when soil moisture conditions can be relatively well-controlled, our findings should be confirmed in multi-year studies in farmers' fields. These results suggest that in flooded rice systems receiving organic amendments or organic management, the e-AWD water regime can achieve multiple environmental and food safety objectives without compromising yield.

## 1. Introduction

A major challenge facing global agriculture in the 21st century is the provision of sufficient, healthy food for the world's growing population, while concurrently reducing its environmental footprint with the challenges of water scarcity, extensive degradation of soil and climate change (Foley et al., 2011). Rice (*Oryza sativa* L) is the staple food of more than half the population of the world (IRRI, 2013) and is grown on over 163 million ha of cropland (FAOSTAT, 2014). It is also the main irrigated crop worldwide, covering 29 % of the planet's total irrigated crop area and almost 50 % of the irrigated cereal area (FAO, 2018). Estimates suggest that rice production is responsible for the consumption of more than 45 % of all freshwater resources in Asia, and approximately 30 % of the world's irrigation water (Bouman and Tuong, 2001; Bouman, 2007). Furthermore, an analysis of crop yield trends combined with models shows that global rice productivity will need to increase by 8–10 million Mg per year in the next ten years to meet global demand (Seck et al., 2012). This has highlighted the need for sustainable intensification (Godfray et al., 2011), i.e. achieving higher yields while reducing damage to the environment, protecting natural resources and ensuring global food security. Increasing production also implies a greater demand for water. Nevertheless, freshwater resources worldwide face the strain imposed by a rapidly increasing population and associated demands for water for urban and industrial use (Bouman, 2007).

One specific concern is that the increased demand for rice will contribute to human-induced climate change because rice fields are a significant source of global $CH_4$ emissions (Smith et al., 2007). Approximately 30 % and 11 % of global agricultural $CH_4$ and $N_2O$ emissions respectively originate from rice cultivations (US-EPA, 2006; IPCC, 2007). Saturated soil conditions and elevated $CH_4$ emissions have resulted in the global warming potential (GWP) of GHG emissions from rice on a per ha basis already being four times higher than for maize and wheat (Linquist et al., 2012). Moreover, as a result of warmer temperatures and higher atmospheric $CO_2$ concentrations, $CH_4$ emissions from rice fields in future may increase by up to 58 % (Van Groenigen et al., 2013). The $CH_4$ load in the atmosphere is therefore projected to increase from about 300 Mt yr$^{-1}$ in 1980 to 750 Mt yr$^{-1}$ in 2100 if no action to mitigate the situation is taken (Van Groenigen et al., 2013).

A highly promising option to decrease water use and reduce GHG emissions without jeopardising rice production is alternative water management such as alternate wetting and drying (AWD). This approach allows fields to dry for a certain number of days and only applying irrigation water when plants show signs of water shortage or when a certain threshold of soil water potential is reached (Tuong et al., 2005; Bouman, 2007). This strategy has been shown to decrease irrigation water use and reduce GHG emissions while maintaining or improving yields (Richards and Sander, 2014). Although AWD substantially reduces $CH_4$ emissions (Qin et al., 2010; Pandey et al., 2014; Linquist et al., 2015; Xu et al., 2015), it can lead to higher $N_2O$ emissions, resulting in a trade-off (Yang et al., 2012; Ahn et al., 2014). Furthermore, the reported effect of AWD on grain yield is very variable. Some studies have shown that AWD can bring about yield reductions of up to 70 % compared to the flooded control (Bouman and Toung, 2001; Linquist et al., 2015; Xu et al., 2015), while many others have shown no change in yield at all (Belder et al., 2004; Dong et al., 2012; Yao et al., 2012; Pandey et al., 2014) or even an increase (Liu et al., 2013a,b). The variable yield response to AWD is not unexpected given the extensive range of water management regimes classified as AWD. These AWD practices can broadly be classified into two categories: i) mild AWD and ii) severe AWD (Carrijo et al., 2017). The mild AWD practice uses field water level (FWL) ≤ 15 cm or soil water potential (SWP) ≥ −20 kPa as AWD thresholds, and can reduce water use by 7–25 % and $CH_4$ emissions by up to 48 % with no effect on grain yield. Although severe AWD (SWP < −20 kPa) can reduce water usage by up to by 33 % and methane emissions by up to 90 %, it has often resulted in a yield penalty, which is a serious concern for farmers and for the sustainability of the production systems. This has highlighted how crucial the extent and timing of AWD are for grain yields (Boonjung and Fukai, 1996). The challenge therefore lies in modifying mild AWD or having an optimal combination of both practices, which can strongly mitigate greenhouse gas emissions in the same way as severe AWD while simultaneously increasing grain yield. Although the severity of AWD has previously been investigated, its timing has not been extensively studied and could provide a potential solution to this aspect. A recent laboratory study involving the timing and duration of drainage by Islam et al. (2018) showed that long early-season drainage (LED) can result in a remarkable emission reduction by stabilising the dissolved organic carbon (DOC) early in the season. The present study's proposed early AWD (e-AWD) water regime combines this knowledge of LED with the mild AWD practice. This is also related to the "safe AWD" practice promoted by the International Rice Research Institute (IRRI), which uses a FWL of 15 cm as the threshold for rewetting so as to avoid a penalty in yield (Bouman, 2007). This e-AWD has the potential to be the optimum setting for AWD practice because its LED component may stabilise reactive C (from organic fertilisers and soil) early in the season, leading to less substrate for methanogens and increasing redox potential (Eh), all of which will suppress methanogenic activity and favour methanotrophs instead so that $CH_4$ is oxidised, while a safe AWD threshold will ensure it reduces water use without negatively affecting yield. From the perspectives of farmers, saving water without reducing grain yield is the ultimate focus for the adoption of such a practice, while a reduction in GHG emissions is considered a welcome side benefit. Incorporating early-season drainage with one mid-season drainage episode can also reduce $CH_4$ emissions, but the water saving of such practice is minimal. Tariq et al. (2017) tested early + mid-season drainage in Vietnam, but the early drainage started at 15 DAT, which may not be early enough for systems in which large amount of organic fertilisers are applied, e.g. organic production, because a considerable amount of methane can be emitted during this period. Moreover, a simple process of this kind with two episodes of drainage has a limited effect on yield, but a practice such as AWD with multiple drainage episodes throughout the season that saves more water may have adverse effects on grain yield. Therefore, incorporating early-drainage with proven safe-AWD practice could be the way forward in that it reduces significant water use and methane emissions simultaneously without affecting grain yield. This is an area that merits further exploration.

In addition to water management, fertilisation also plays a key role in both yield and GHG emissions (especially $N_2O$) (Mueller et al., 2013). Greater (over)use of mineral fertilisers has resulted in the widespread degradation of natural resources and disturbance of global nutrient cycles (Robertson and Vitousek, 2009; Schlesinger, 2009; Hoekstra and Mekonnen, 2012). Combined with increasing costs of mineral fertilisers, there is increasing interest in the application of organic fertilisers such as manures, composts and digestates to agricultural soil. Application of these organic fertilisers has several advantages such as reducing costs, contributing to climate change mitigation through C sequestration (Diacono and Montemurro, 2010), stimulating soil life, improving soil structure and fertility, and simultaneously helping to tackle waste management issues (Tirado et al., 2010). There is greater interest in rice production in accordance with organic farming principles, and already a growing market in Asia, Europe and the USA for rice produced under organic standards and certification schemes, i.e. without synthetic mineral fertilisers. Owing to its higher unit price and perceived health benefits, certified organic production is projected to increase rapidly, which means an increasing amount of rice will be produced in future using organic fertilisers.

However, there are a number of drawbacks to using organic fertilisers, which include potential contamination with heavy metals as well as higher GHG emissions (Petersen et al., 2003; Snyder et al., 2009). Heavy metals or metalloids in manure are a concern in organically fertilised rice systems owing to metals being added to animal feed in intensive animal productions (Paradelo et al., 2011a,b; Wang et al., 2013). The main heavy metals in rice systems are arsenic (As), cadmium (Cd) and lead (Pb), and health risks associated with ingestion of As-contaminated rice are of considerable concern in countries in which rice is a daily staple food (Mandal and Suzuki, 2002; Zhu et al., 2008). Due to anaerobic reduced conditions typically associated with flooded rice cultivation, As is reduced from As (V) to more mobile As (III) (Takahashi et al., 2004), leading to higher soil solution As concentrations. This increases its phyto-availability and uptake by rice since the dominant species taken up by rice roots is As (III) (Chen et al., 2005; Ma et al., 2008; Zhao et al., 2010). Previous studies in the USA with mineral fertilisers have demonstrated reduced grain As accumulation with AWD (Linquist et al., 2015), but its effectiveness in organically produced rice with the various forms of manure applied is still unclear. So far, the common practice in rice systems has been to apply mineral N fertilisers with occasional application of rice straw as an organic amendment, which has also been the focus of previous AWD studies. However, the application of full doses of various forms of organic fertilisers under organic management is relatively new and the impact of AWD in such systems is less clear, which indicates the need for a more comprehensive and systematic study.

The present study was the first to test an e-AWD system combining "LED" and "safe AWD" to maximise the potential of AWD practice for reduction of greenhouse gas emissions, water use and grain As accumulation while maintaining crop yield. The benefits associated with AWD management in rice systems have been investigated separately in various studies, however the objective of the present study was to evaluate the multiple benefits of e-AWD in a single comprehensive investigation to facilitate better understanding of the processes and interactions involved. It was hypothesised that an e-AWD water management regime can maintain grain yield with the application of similar N rates of various organic fertilisers alone or in combination, while decreasing total As levels in the grain and GWPs from rice production.

## 2. Materials and methods

### 2.1. Site characteristics and experimental setup

A field experiment was set up at the Experimental Station of the International Rice Research Institute in Los Baños in the Philippines (14°09′42.3″N 121°15′48.0″E). It is located at an elevation 21 m above mean sea level, has average yearly rainfall of 2115 mm (2000–2016) and an average annual mean air temperature of 27.4 ± 0.36 °C. The study was conducted during the 2016 dry season (DS). The soil has been classified as Aquandic Epiaquoll (Soil Survey Staff, 1994) with a 62 % a clay content. The soil properties include pH = 6.54, total C (g kg$^{-1}$) = 21.6, total N (g kg$^{-1}$) = 1.9, Olsen P (mg kg$^{-1}$) = 18, exchangeable K (c mol kg$^{-1}$) = 1.9 and CEC (c mol kg$^{-1}$) = 39. The study site was historically cropped with paddy rice. The land was left as fallow and no artificial chemicals were applied in the three years leading up to the experiment. The experiment comprises a split-plot design with three replications. The main plots had two water treatments, while subplots consisted of nine fertiliser treatments. The treatments are given in Table 1.

An inbred rice variety (NSIC Rc18) was used with 20 × 20 cm spacing in plots of 4 × 4 m. At the start of the rice growing season, land was initially prepared with dry cultivation using a plough, followed by land soaking, ploughing, harrowing and levelling using a two-wheel tractor with wooden planks, which incorporated the respective organic fertiliser amounts in all the management treatments. On 30 January 2016, fourteen-day-old rice seedlings were transplanted manually with a spacing of 20 cm x 20 cm and were manually harvested at 102 days after transplanting (DAT). Cultural and mechanical practices of integrated pest management (IPM) were followed in the experimental plots. These included the ploughing the field in summer, healthy seed selection, timely planting, the raising of a healthy nursery, the removal of weeds from the field, regular field monitoring and pest-infested plant parts removal and destruction, the clipping of rice seedling tips and the collection of pest egg masses and larvae. To control snails, in the early morning and afternoon when snails are most active, they were hand-picked, and egg masses were crushed. Bamboo stakes were placed in each plot to provide sites for egg laying that allowed easy collection of snail eggs for destruction. A wire screen was also placed on the main irrigation water inlet and outlet to prevent snail entry. In order to stop rats and other pests from entering the experimental plots, the whole experimental site was fenced with a tough plastic sheet positioned up to 1 m belowground and 50 cm aboveground. No insecticides, pesticides or herbicides were applied in any of the treatments at any point in the experiment. Plots were separated by a 1-m deep plastic barrier to prevent contamination and lateral movement of nutrients. Organic fertilisers were applied five days before transplanting, with the practice closely following organic farming principles. In combination treatment plots, organic fertilisers were similarly applied five days before transplanting. Mineral N fertilisers were applied following common practice in three split applications of 30 %, 35 % and 35 % at 10, 32 and 46 DAT respectively. The soil was kept saturated with 2−3 cm water in the early vegetative stage from 0 to 3 DAT to allow the seedlings to recover from the transplanting shock. Subsequently, in the conventional (CF) water regime, a 5-cm water depth was maintained in the field until 10 days before harvest (92 DAT). In order to monitor the depth of the water table in the field and schedule irrigation in e-AWD plots, a number of perforated PVC pipes (10 cm diameter, 30 cm long, 5 mm slot size at a spacing of 5*5 cm) were installed 10 cm above the soil surface and 20 cm below the soil surface. In the e-AWD water regime, the first drainage was implemented as early as 4 DAT for seven days (LED) (much earlier than 20 DAT which is common in safe AWD practices); afterwards irrigation was scheduled when the soil water level reached a depth of 15 cm and the wetting and drying cycles continued throughout the season (drainage usually stops at the flowering stage in the safe AWD practice). In both water regimes, the plots were drained ten days before harvest to allow the grains to fully ripen and for the fields to dry for easy harvesting.

### 2.2. Gas sampling, analysis and calculation

The collection of gas samples was carried out on 29 occasions between -7 and 116 DAT, which included the land preparation, growth and fallow periods. Samples were always taken in the daytime between 8.00 am and 11.30 am. After each N application, gas measurements were taken daily for five consecutive days. Every single chamber was anchored by a stainless-steel metal base (40 cm length x 22 cm x width 12 cm height) which was inserted into the soil at about 10 cm depth. The chamber contained two rice hills. The gas flux chambers were made of Plexiglas with measurement of 40 cm length x 22 cm width, were used at variable heights (11, 42, and 81 cm) to adapt the height of the growing plants inside the chamber (Sander et al., 2014). Each chamber comprises of a vent to allow pressure equilibration, two fans, a thermometer, and a gas sampling port. Gas samples were collected by a 60-mL syringe fitted with a stopcock at 0, 10, 20 and 30 min after the chamber was closed. The gas samples were then immediately injected into an evacuated 30-mL vial, and the concentrations of $CH_4$ and $N_2O$ analysed using a gas chromatograph (SRI GC-8610C) equipped with separate detectors for CH4 and $N_2O$. $N_2O$ was determined by electron capture detector (ECD) operated at 350 °C and $CH_4$ was determined by flame ionisation detector (FID) operated at 300 °C. The $CH_4$ and $N_2O$ fluxes were calculated in accordance with Smith and Conen (2004) and Vu et al. (2015). To calculate GWP in $CO_2$ equivalents based on a 100-year time horizon, seasonal $CH_4$ emissions were multiplied by a factor of 34 and seasonal N$_2$O emissions by a factor of 298 (Hou et al., 2012; IPCC, 2013; Wang et al., 2015). The yield-scaled GHG emissions were calculated by dividing CO$_2$-equivalents emissions by rice yield according to Van Groenigen et al. (2010).

**Table 1**  
List of fertiliser treatments.

| No. | Fertiliser management | Composition | Total N/ha |
|---|---|---|---|
| | *Organic fertiliser only* | | |
| 1 | FYM | 100 % farmyard manure | 160 kg N |
| 2 | Compost | 100 % compost | 160 kg N |
| 3 | Digestate | 100 % biogas digestate | 160 kg N |
| | *Organic + mineral fertiliser* | | |
| 4 | FYM + Urea | 50 % farmyard manure + 50 % urea | 160 kg N |
| 5 | Compost + Urea | 50 % compost + 50 % urea | 160 kg N |
| 6 | Digestate + Urea | 50 % biogas digestate + 50 % urea | 160 kg N |
| 7 | Compost + AS | 50 % compost + 50 % ammonium sulfate | 160 kg N |
| | *Control treatments* | | |
| 8 | Urea | 100 % urea | 160 kg N |
| 9 | Unfertilised control | No fertiliser | 0 kg N |

### 2.3. Soil and plant analyses

Soil samples were collected 11 times during the rice-growing season and analysed for ammonium, nitrate and DOC. Samples of soil were collected from all three replicates of the 18 treatments, with five sub-samples taken from the top 20 cm at one at the middle and four corners of each field plot. These sub-samples were then bulked into a composite sample representing the true field replicate (Gómez-Muñoz et al., 2017). Collected soil samples were stored at −20 °C until further analysis. Soil NH$_4^+$ and NO$_3^-$ concentrations were analysed after extraction with a 1 M KCl solution by flow injection analysis (FIAstar 5000 flow injection analyser (Foss Analytical, Hillerød, Denmark)) according to Gómez-Muñoz et al. (2017). DOC was measured after soil extraction with ultra-pure water (UPW), as described by Straathof et al. (2014), and analysed on a Shimadzu TOC-VCPN (Kyoto, Japan) for DOC. Microbial biomass carbon was determined by using the chloroform fumigation extraction method modified for flooded rice soils (Vance et al., 1987; Inubushi et al., 1991). Soil pH was analysed in H$_2$O suspensions (1:5 w/v). CEC (Cation exchange capacity) of soil was measured with 0.1 M BaCl$_2$ solution with ICP-AES (Thermo iCAP 6500 DV; Thermo Fisher Scientific). Total Nitrogen of soil samples was determined by first digesting the organic compounds with potassium persulfate at pH = 4 followed by UV-digestion by means of sodium borate using a segmented flow analyzer (SFA). Total organic carbon was measured using 0.01 M CaCl$_2$ extracts using SFA (NEN-EN 1484, 1997; Houba et al., 2000).

To measure root biomass, samples were collected at harvest (102 DAT) using a Monolith sampler (20 × 20 cm, 45 cm depth, as described by Henry et al. (2011)). Roots were washed based on the "Goetingen method" as described by Böhm (1979) and root samples were collected in a coin envelope, oven-dried and weighed. To determine root length, roots were arranged and floated on shallow water in a glass tray (30 × 30 cm), scanned and analysed by WinRHIZO Root Analyzer System (Regent Instruments Inc., Quebec, Canada). Each plot contained platinum electrodes that were permanently installed and placed in pairs at 0.1 m depth for redox measurement. Soil redox potential was measured 23 times during the rice-growing season (always between 9.00 am and 11.00 am), compared to an Ag/AgCl reference electrode at a soil temperature of 30 ± 2 °C. Subsequently, the readings were converted to the standard hydrogen electrode reference at 25 °C by addition of 197 mV (PCRA, 2007). The plants attained physiological maturity at 102 DAT with more than 80 % ripe grains and the aboveground plant biomass and grains were then harvested, oven-dried and weighed. Grain samples were prepared and heavy metals were analysed by ICP-MS (Agilent 7900, USA) according to the standard protocol (D'Ilio et al., 2002; Spanu et al., 2012; Linquist et al., 2015; Pan et al., 2016). Briefly, a decontaminated agate-ball mill was used to ground the rice grains. The ground rice flour thus obtained was subsequently digested by acid-assisted microwave (MW) irradiation using a commercially available oven (MLS-1200 Mega, FKV, Bergamo, Italy). The rice grain sample of 0.5 g was placed in a PTFE vessel and a mixture added of 5 ml 65 % HNO$_3$ Suprapur (Merck, Darmstadt, Germany), 1 ml 30 % H$_2$O$_2$ Suprapur (Merck) and 1 ml high purity deionised water (EASY pure UV, Barnsteady Thermolyne, Dubuque, IA, USA) (18 MO). After being closed, the vessel underwent the relevant digestion cycle: 1 min 40 s at 250 W, 2 min at 0 W, 6 min at 250 W, and then 8 min at 400 W. The vessel temperature at the end of the digestion cycle was typically no higher than 120 °C. Then the vessel was kept under the ice for about 2 h for cooling. After opening the vessel, the content was diluted with up to 20 cm$^3$ of water and then a 0.45 μm polypropylene filter was used to filter the solution. Single-element calibrants were prepared from 1000 mg ly$^1$ stock solutions of As, Cd, Co, Cr, Cu, Fe, Mn, Pb, V and Zn in 2 % HNO$_3$ (Spex Industry, Edison, NJ, USA) by dilution with high-purity deionised water. Horwitz's theory (Horwitz, 1982) was used to successfully verify the acceptability of the precision values. To evaluate the trueness, repeated analyses on three CRM rice flours at different total As concentrations, ranging between 290 ± 30 μg kg$^{-1}$ (NIST SRM 1568a) and 49 ± 4 μg kg$^{-1}$ (IRMM 804) was utilized. The obtained recovery values ranged between 96 % and 97 %.

### 2.4. Carbon payment

For the carbon payment, the June 2016 price of mitigation of US$ 12.75 Mg$^{-1}$ of CO$_2$ equivalent (eq) from the European Carbon Futures Market on the European Energy Exchange (EEX) was used. To calculate the profit for each alternate irrigation regime, the simplified version from Nalley et al. (2015) was adopted (Table 3). The profit for irrigation regime j including C payment used can be written as:

$$\pi j = (PY_j) - C_j + (P_{CO_2} X_{CO_2 j}) \qquad (1)$$

where P is the average price for a kilogram of rice in 2016, $Y_j$ is the yield under irrigation regime j, $C_j$ is the other costs of production (seed, fertiliser, labour etc.) for each irrigation regime j, which were assumed to be fixed because the same variety was used. It was therefore dropped from the relative comparison. This meant that profits in absolute terms estimated in Eq. [1] were unbiased. $P_{CO_2}$ is the price of Mg$^{-1}$ of CO$_2$ and $X_{CO_2 j}$ is the amount of total CO$_2$ eq GHG (CH$_4$ and N$_2$O) reduction from irrigation regime j compared with traditional flooding. Compared to the carbon payment from the reduction of methane, the payment for diesel use reduction for irrigation is relatively small, and are not applicable in areas under gravity-driven irrigation systems. These were therefore not considered in this calculation.

### 2.5. Statistical analysis

In this study Statistical Analysis System (SAS) software version 9.4 (SAS Institute Inc., USA) was used for statistical analysis. The normality, independence and homogeneity of variance of the dataset were examined and all the data met the assumptions without transformation. PROC MIXED in SAS software was used for analysis of variance with the general linear model (GLM) procedure on the growing season's cumulative CH$_4$ emissions, N$_2$O emissions, yield, and grain As, Pb and Cd. A two-way ANOVA was used with two water management regimes (CF and e-AWD) and fertiliser treatments were included as fixed effects and the block and block-by-treatment interaction was included as random effects to test their impact on CH$_4$, N$_2$O emissions, GWP and yield-scaled GWP. Tukey's HSD test was utilized to determine the significant difference at the 95 % level ($P < 0.05$).

Structural equation modelling (SEM) was performed using the LAVAAN R package (Rosseel, 2012) to identify the direct and indirect controls of plant and soil variables on CH$_4$ and N$_2$O emissions and thereby improve overall mechanistic understanding. *A priori* models were created on the basis of the hypotheses of variables affecting CH$_4$ and N$_2$O emissions. Prior to SEM analyses, the units of the predictor and dependent parameters were adjusted to obtain comparable parameter variances. Model modification indices were applied to remove non-significant relationships following a stepwise procedure (Abalos et al., 2016). The effect of these removals was tested using the Akaike information criterion (AIC) and model fit was tested using a likelihood ratio test (De Vries and Bardgett, 2016). The model fit was assessed using the X$^2$ goodness of fit statistic (P-values > 0.05 indicates a statistically significant model fit), the root mean square error of approximation value (RMSEA), the comparative fit index (CFI), the standardised root mean square residual (SRMR), the Bayesian information criterion (BIC) and the Akaike information criterion (AIC) (Grace, 2006; Kline, 2011).

**Table 2**

Rice grain yields and grain total arsenic, lead and cadmium concentrations of polished white rice in e-AWD vs. the CF water regime in a) organic systems b) combination systems.

(a)

| Water treatment | Organic fertiliser treatment | Rice grain yields* (Mg ha$^{-1}$) | Rice grain concentration* (µg kg$^{-1}$) | | |
|---|---|---|---|---|---|
| | | | Arsenic (As) | Lead (Pb) | Cadmium (Cd) |
| CF regime | FYM | 5.5$^c$ | 460$^a$ | 110$^a$ | 40$^a$ |
| | Compost | 5.9$^{bc}$ | 320$^b$ | 90$^b$ | 30$^a$ |
| | Digestate | 6.3$^a$ | 450$^a$ | 60$^c$ | 30$^a$ |
| | Unfertilised control | 3.3$^d$ | 180$^c$ | 30$^d$ | 20$^b$ |
| e-AWD regime | FYM | 5.8$^{bc}$ | 160$^d$ | 30$^d$ | 30$^a$ |
| | Compost | 6.5$^a$ | 180$^c$ | 20$^d$ | 20$^b$ |
| | Digestate | 6.6$^a$ | 190$^c$ | 20$^d$ | 20$^b$ |
| | Unfertilised control | 2.8$^e$ | 150$^d$ | 20$^d$ | 20$^b$ |

(b)

| Water treatment | Combination fertiliser treatment | Rice grain yields* (Mg ha$^{-1}$) | Rice grain concentration* (µg kg$^{-1}$) | | |
|---|---|---|---|---|---|
| | | | Arsenic (As) | Lead (Pb) | Cadmium (Cd) |
| CF regime | FYM + Urea | 5.7$^{cd}$ | 240$^a$ | 70$^a$ | 30$^a$ |
| | Compost + Urea | 7.1$^{ab}$ | 200$^b$ | 50$^b$ | 20$^a$ |
| | Digestate + Urea | 6.1$^{cd}$ | 230$^a$ | 40$^b$ | 30$^a$ |
| | Compost + AS | 6.9$^{bc}$ | 200$^b$ | 45$^b$ | 20$^a$ |
| | Urea (control) | 6.9$^{bc}$ | 180$^c$ | 30$^c$ | 20$^a$ |
| e-AWD regime | FYM + Urea | 6.4$^{bc}$ | 150$^c$ | 30$^c$ | 30$^a$ |
| | Compost + Urea | 7.5$^a$ | 130$^d$ | 20$^c$ | 30$^a$ |
| | Digestate + Urea | 7.1$^{ab}$ | 140$^d$ | 20$^c$ | 30$^a$ |
| | Compost + AS | 7.5$^a$ | 130$^d$ | 20$^c$ | 25$^a$ |
| | Urea (control) | 6.7$^{bc}$ | 130$^d$ | 20$^c$ | 30$^a$ |

* Within the column, the values with different letters are significantly different at the p < 0.05 level.

## 3. Results

### 3.1. Rice yield and grain heavy metal content

In all the treatments in which organic fertilisers were applied (either with or without artificial fertiliser), e-AWD produced higher yields compared to CF (Table 2), although this increase was only significant (p < 0.05) for the Compost, Digestate + Urea and Compost + AS treatments. For the e-AWD treatments receiving organic fertiliser, the highest yields were found for two combination treatments (Compost + Urea and Compost + AS) at 7.5 Mg$^{-1}$, while the lowest yields were for the FYM treatment (5.8 Mg ha$^{-1}$).

Total grain arsenic concentrations ranged from 180 to 460 µg kg$^{-1}$ in the various organic fertiliser treatments (Table 2a) and from 200 to 240 µg kg$^{-1}$ in the combination treatments (Table 2b) under conventional CF. Overall, arsenic levels were reduced by 57 % (p < 0.05) by the e-AWD water regime in organic fertiliser treatments, while in combination treatments the reduction of arsenic levels was lower, averaging just 37 %. A similar trend was observed for lead; on average the e-AWD water regime resulted in a reduction in grain lead concentrations in organic and combination treatments of 72 % and 56 % (p < 0.05) respectively. The e-AWD water regime reduced grain cadmium levels by 33 % in both compost and digestate treatment. However, no significant differences in Cd concentrations in rice grain were found between the fertiliser and water treatments in combination treatments.

### 3.2. GHG fluxes

Fig. 1 shows the temporal dynamics of the CH$_4$ fluxes for all treatments and Table 3 gives the cumulative emissions. Seasonal CH$_4$ emissions from the FYM, Compost and Digestate treatments were reduced in the e-AWD water regime by 69, 76 and 85 % respectively. Overall, e-AWD reduced CH$_4$ emissions by 77 % ($p < 0.01$). In the treatments with organic fertilisers only, both the FYM and Digestate treatments showed a higher initial CH$_4$ flux after transplanting compared to the Compost treatment (Fig. 1). FYM had the highest peak of the three organic fertiliser treatments at 51 mg m$^{-2}$ h$^{-2}$, which was reached three weeks after transplanting. In the e-AWD water regime, emissions from FYM and digestate were considerably reduced. Of the three organic fertiliser treatments, compost showed the lowest CH$_4$ emissions and peaks in both water treatments. Methane emissions from combined mineral and organic fertiliser treatments were lower than treatments receiving only organic fertiliser in both water regimes. This was due in particular to a much lower first peak than the conventional CF and no second peak. Methane emissions from the FYM + Urea, Comp + Urea, Compost + AS, Digestate + Urea treatments were reduced by 74, 72, 51 and 71 % respectively in the e-AWD water regime.

The largest proportion of N$_2$O emissions was limited to peaks related to fertiliser application and soil drainage events as part of the e-AWD water management regime (Fig. 2). During the rice-growing period two to four N$_2$O emission peaks were found, depending on the fertiliser treatments. Cumulative N$_2$O emissions were significantly affected by both the fertiliser and water regimes (Table 3). In comparison with the overall low emissions under CF, e-AWD water treatments resulted in a 23–305 % increase for treatments receiving organic fertiliser only, and a 33–292 % increase for the treatments receiving combined organic and mineral fertiliser applications (Table 3). The seasonal total N$_2$O emissions ranged from 0.45 to 1.85 kg N$_2$O ha$^{-1}$ across organic and combination fertiliser treatments.

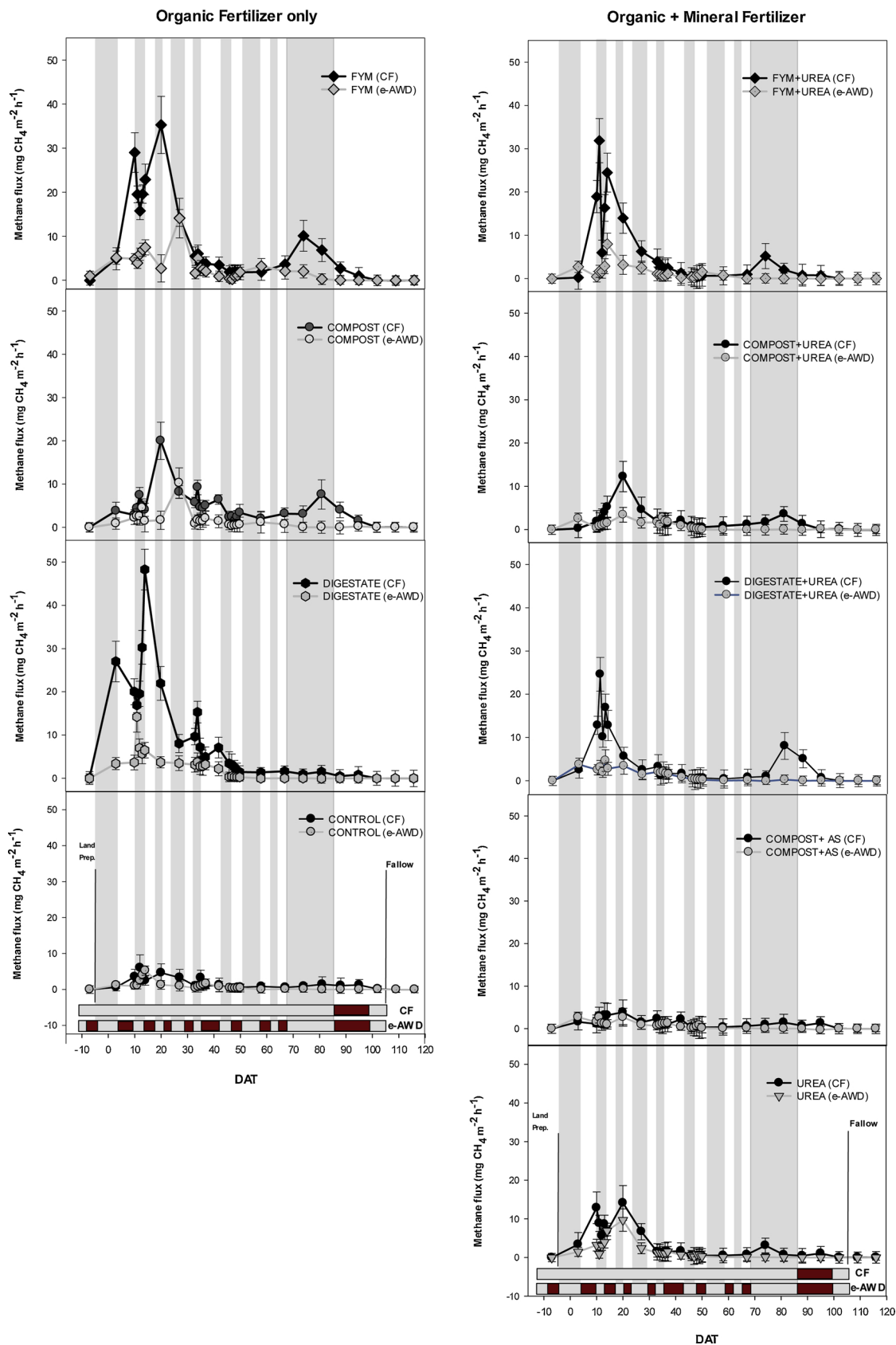

**Fig. 1.** Dynamic temporal patterns of the methane flux of e-AWD vs. the CF water regime as affected by organic fertilisers only or by organic and mineral fertiliser combinations. The approximate time the soil was flooded or saturated in the e-AWD water regime is represented by the shaded bars. Error bars indicate 1 S.E.M. (n = 3).

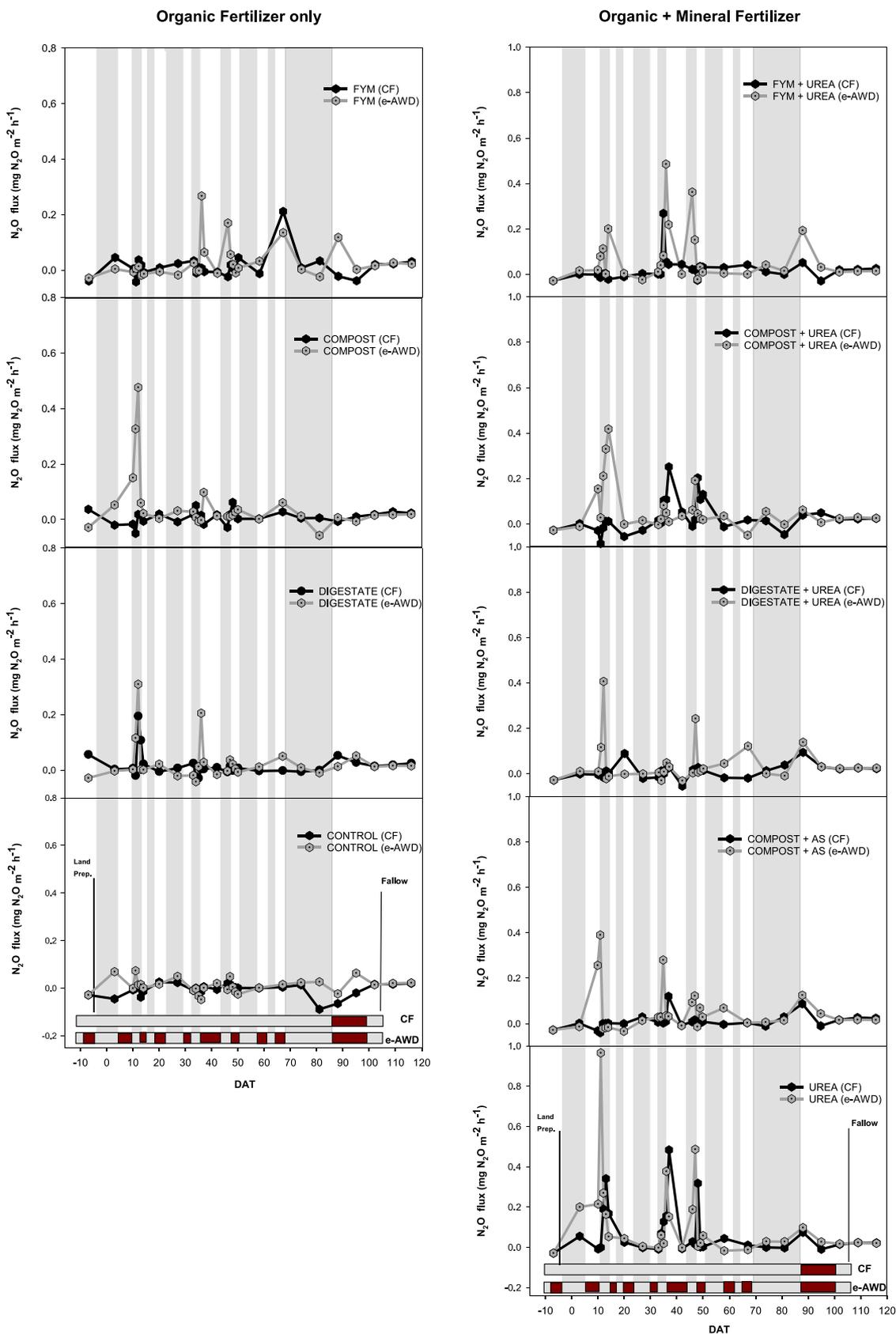

**Fig. 2.** Nitrous oxide flux of e-AWD vs. the CF water regime as affected by organic fertilisers only or organic and mineral fertiliser combinations. Error bars are omitted for improved clarity. The shaded bars represent the approximate time the soil was flooded or saturated in the e-AWD water regime.

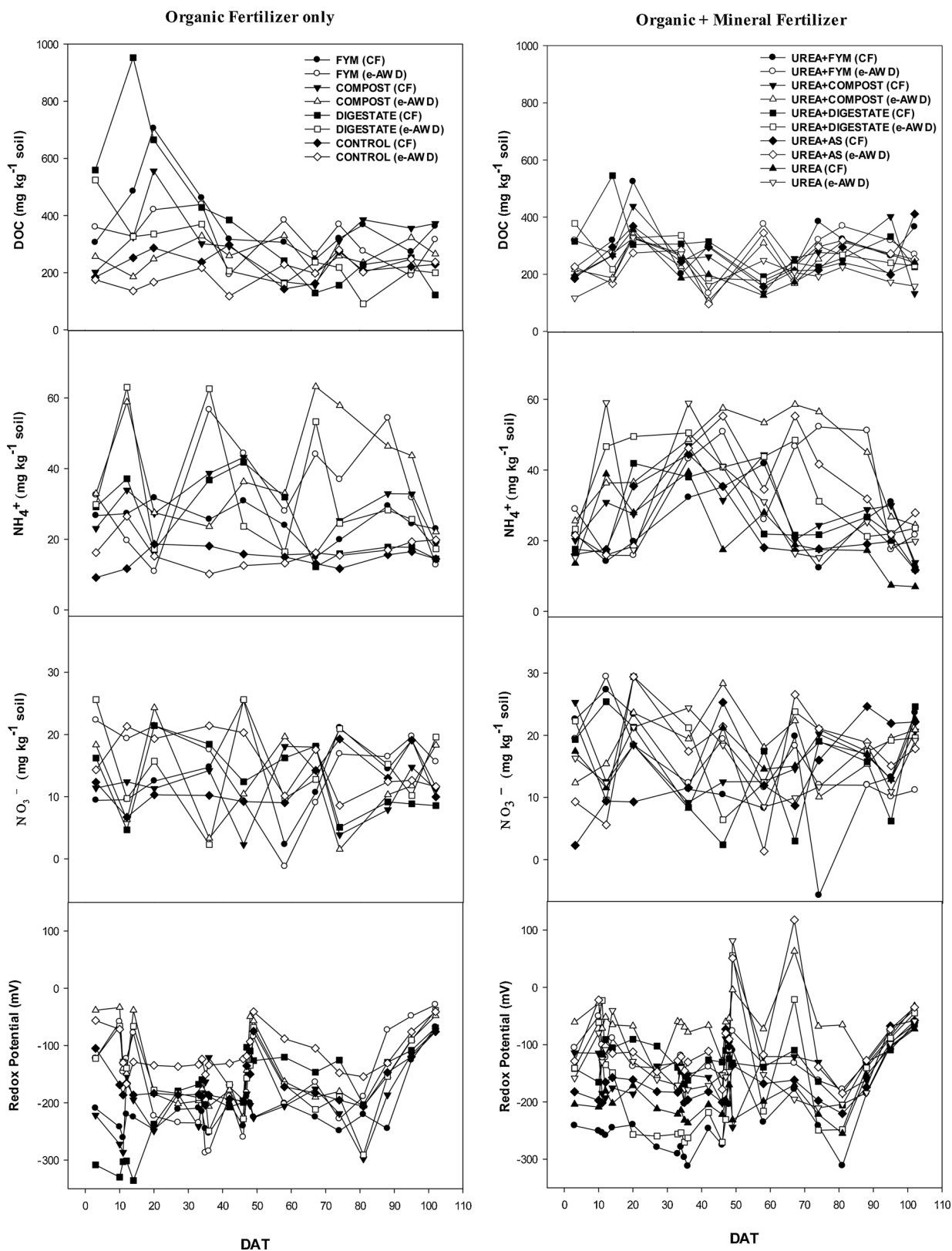

**Fig. 3.** Dissolved organic carbon (DOC), extractable mineral N ($NH_4^+$ and $NO_3^-$) and redox potential as affected by the e-AWD water regime and fertiliser treatments.

### 3.3. Soil and root parameters

Patterns of DOC differed clearly between the CF and e-AWD water regimes, most notably by a sharp decrease (21–47 %) after early-season drainage in the e-AWD regime (Fig. 3).

CF also resulted in a much higher soil $NH_4^+$ content during the vegetative and ripening stage of rice growth (Fig. 3; S1), whereas higher soil $NH_{4+}$ was observed during the vegetative and reproductive stage in the e-AWD water regime, where $NH_4^+$ availability was 33–45 % higher from different organic treatments and 37–48 % higher from the organic + mineral fertiliser combination treatments relative to the control CF. $NO_3^-$ availability increased through the aerobic cycles implemented as part of the e-AWD water regime by 10–27 % for organic treatments and by 5–24 % from the combination fertiliser treatments.

The redox potential (Eh) under CF ranged between -68 and −329 mV for treatments receiving organic fertilisers and from -61 to −255 mV for mineral and organic combination fertiliser treatments respectively (Fig. 3). Under the e-AWD water regime, these ranges were -38 to −287 mV and +117 to −318 mV respectively. Generally, the drainage episodes in e-AWD water regime increased the soil redox potential whereas re-flooding decreased the redox potential of soil. The $CH_4$ emission peaks were associated with periods of very low redox potential. On the other hand, higher $N_2O$ emissions were associated with high soil redox potential.

Relative to CF, e-AWD resulted in lower microbial biomass carbon (MBC) for the FYM and Digestate treatments and a 46 % increase for the Compost treatment (Fig. 4). In treatments receiving combined fertiliser applications, e-AWD increased MBC by 58–312 %. Generally, root biomass and root length increased due to e-AWD in all treatments. Relative to the CF water regime, root biomass and root length increased in organic rice treatments by 9–72 % and 28–41 %, while in combination rice treatments under e-AWD water management it increased by 10–41 % and 20–33 %, respectively.

### 3.4. Structural equation model (SEM)

Fig. 5 shows the outcome of the structural equation model (SEM). This conceptual model fitted well with the measured data in this study ($\chi^2$ = 0.766-0.854, p = 0.48-0.80, CFI = 1, RMSEA = 0.86-0.87, AIC = 766–992). Modification indices (mi) values were low, indicating that this model could not be improved more by adding further relationships. For $CH_4$ emission, the SEM revealed that dissolved organic carbon (DOC), root biomass, redox and MBC regulate the $CH_4$ emissions. Root biomass was found to play a central role by directly and/or indirectly influencing all the factors involved, and had a strong positive influence on grain yield and DOC. DOC had the strongest significant relationship with the $CH_4$ emissions of all the parameters. However, the relationship between yield and $CH_4$ emissions was not significant. The effect of soil redox potential was also highly significant, but its contribution was low compared to the other three factors. For $N_2O$ emission, soil $NH_4^+$, $NO_3^-$ and redox potential played a central role. The $NH_4^+$ had the strongest significant relationship with the $N_2O$ emissions of all the parameters. Root biomass was found to significantly affect the $NH_4^+$ and DOC but unlike the $CH_4$ emission, the relationship between DOC and $N_2O$ emissions was not significant, but it did improve the fit of the model.

### 3.5. Area-scaled and yield-scaled GWP, carbon payment

Although the relative increase in $N_2O$ emissions was considerable,

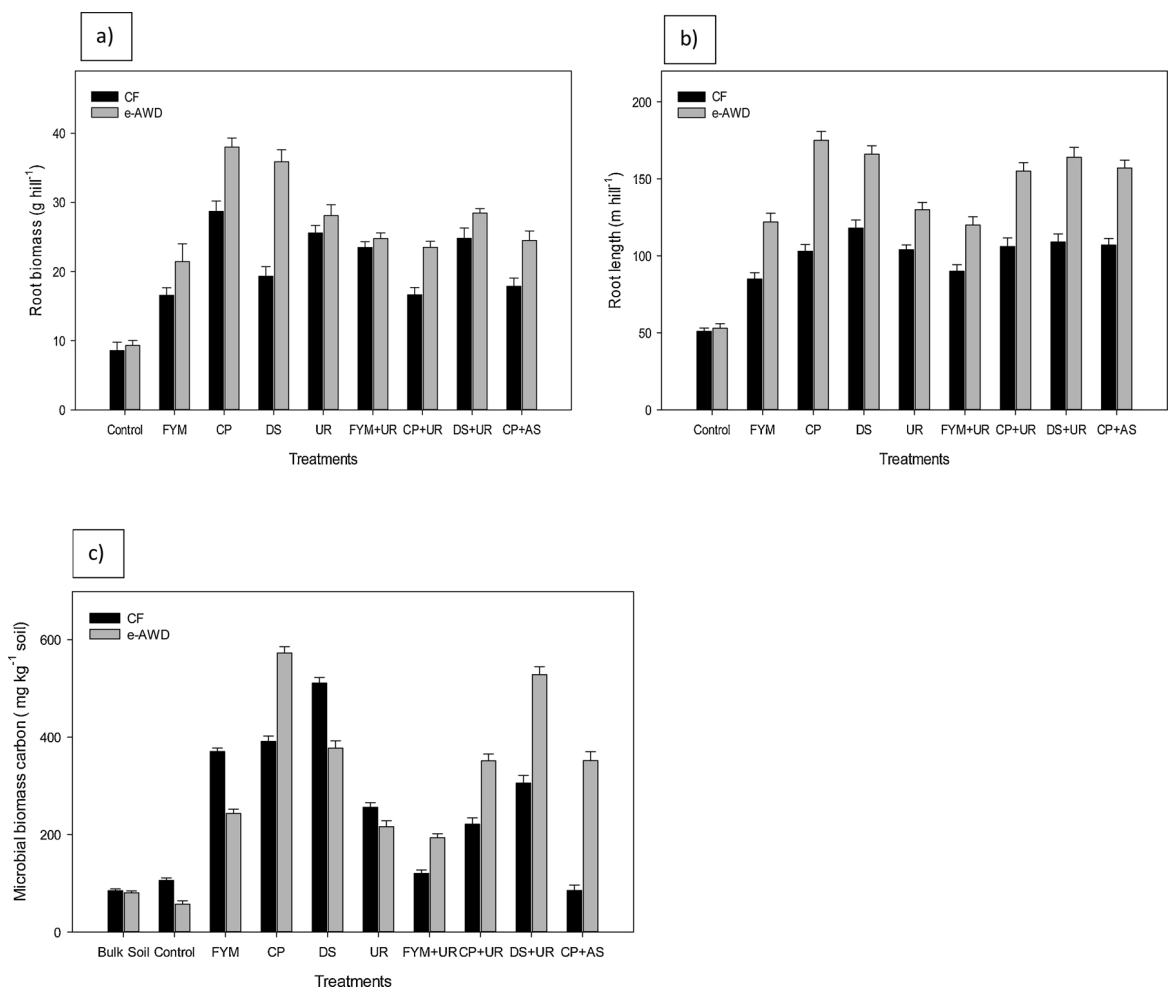

**Fig. 4.** Root biomass (a), root length (b) and microbial biomass carbon (c) as affected by e-AWD water regime and fertiliser treatments. FYM = farmyard manure, CP = compost, DS = digestate, UR = Urea, AS = Ammonium sulphate.

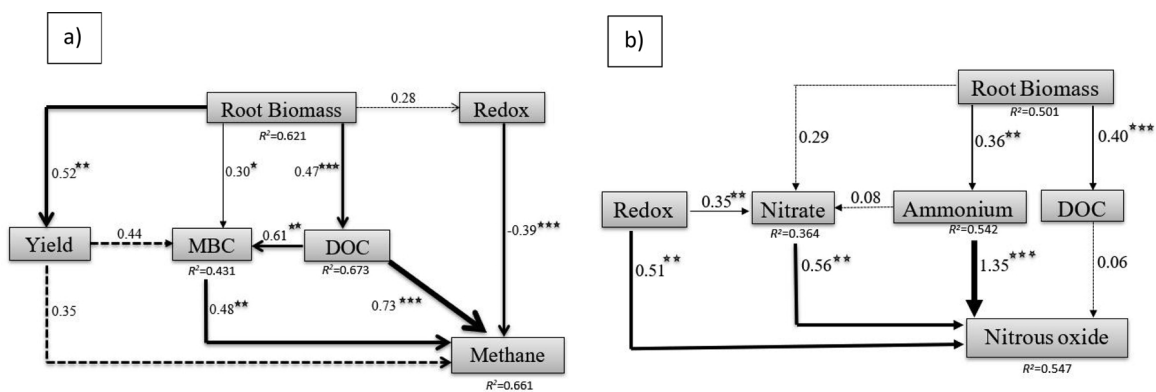

**Fig. 5.** Structural equation model (SEM) to explain: a) CH$_4$ emission mediated by soil parameters and plant traits in rice ecosystems, and b) N$_2$O emission mediated by soil parameters and plant traits in rice ecosystems. The weight of the arrows indicates the strength of the causal relationship. Dashed arrows represent non-significant relationships and continuous arrows represent significant relationships (*p < .05; **p < .01; ***p < .001). Numbers next to the arrows represent the (positive or negative) path coefficients. R$^2$ values denote the amount of variance explained by the model for the response variables. The fit of model a was good ($\chi^2$ = 0.854, p = 0.484; comparative fit index = 1.000; root mean square error of approximation < 0.05, p = 0.862, Akaike information criterion = 766.8). The fit of model b was good ($\chi^2$ = 0.784, p = 0.835; comparative fit index = 1.000; root mean square error of approximation < 0.05, p = 0.876, Akaike information criterion = 996.2).

area-scaled emissions expressed as GWP were dominated by CH$_4$ emissions, with their contribution ranging from 80 to 98 % in treatments receiving organic fertiliser. Relative to CF, GWP decreased significantly (p < 0.01) in e-AWD by 69, 70 and 83 % in the FYM, Compost and Digestate treatments respectively (Table 3). This reduction was less in treatments receiving combinations of fertilisers, where it ranged from 29 to 63 %.

The FYM treatment under the CF water regime showed the highest yield-scaled GWP (1.6 kg CO$_2$ equiv. kg$^{-1}$ rice grain), and was significantly (p < 0.01) higher than the other fertiliser treatments in both water regime

(Table 3). Introducing the e-AWD water regime reduced yield-scaled GWP from treatments receiving organic fertiliser alone by 70–84 %, and by 23–65 % for treatments receiving combined fertiliser applications. These results indicate that under the e-AWD water regime, the organically fertilised rice treatments were the largest benefactor of C payment, however overall the most economically attractive options were the combination treatments of compost and digestate with either urea or AS (Table 3).

**Table 3**
Cumulative emissions of CH$_4$ and N$_2$O over the rice-growing season, total CO$_2$ equivalent area-scaled and yield-scaled GWPs over the 100-year time horizon and carbon (C) payment as affected by fertilisers and water management.

| Water treatment | Fertiliser treatment | CH$_4$ emission Kg CH$_4$ ha$^{-1}$ | N$_2$O emission Kg N$_2$O ha$^{-1}$ | GWPs kg CO$_2$ eq. ha$^{-1}$ | Yield-scaled GWPs kg CO$_2$ kg$^{-1}$ rice | Proposed C payment † US$ ha$^{-1}$ | Final price of rice (including C payment) ‡ US$ ha$^{-1}$ |
|---|---|---|---|---|---|---|---|
| CF regime | FYM | 256.1$^a$ | 0.64$^g$ | 8900$^a$ | 1.61$^a$ | – | 2008$^d$ |
| | Compost | 122.5$^c$ | 0.22$^k$ | 4233$^c$ | 0.71$^c$ | – | 2154$^c$ |
| | Digestate | 217.6$^b$ | 0.42$^{ij}$ | 7528$^b$ | 1.20$^b$ | – | 2300$^c$ |
| | FYM + Urea | 115.6$^c$ | 0.47$^{hi}$ | 4073$^c$ | 0.71$^c$ | – | 2081$^d$ |
| | Compost + Urea | 66.4$^{ef}$ | 0.53$^{gh}$ | 2417$^{ef}$ | 0.34$^{fgh}$ | – | 2592$^b$ |
| | Digestate + Urea | 86.7$^d$ | 0.42$^{ij}$ | 3074$^{cd}$ | 0.50$^{de}$ | – | 2227$^c$ |
| | Compost + AS | 31.7$^{gh}$ | 0.38$^j$ | 1193$^{gh}$ | 0.16$^{gh}$ | – | 2519$^b$ |
| | Urea | 74.3$^{ed}$ | 1.19$^c$ | 2887$^{cde}$ | 0.41$^{def}$ | – | 2518$^b$ |
| | Control | 37.6$^{gh}$ | −0.17$^l$ | 1230$^{fgh}$ | 0.37$^{efg}$ | – | 1205$^e$ |
| e-AWD regime | FYM | 75.2$^{ed}$ | 0.79$^f$ | 2796$^{de}$ | 0.49$^{de}$ | 77 | 2194$^c$ |
| | Compost | 29.2$^{ghi}$ | 0.90$^e$ | 1263$^{gh}$ | 0.19$^{fghi}$ | 38 | 2410$^b$ |
| | Digestate | 33.6$^{gf}$ | 0.58$^g$ | 1319$^{fg}$ | 0.20$^{ij}$ | 80 | 2489$^b$ |
| | FYM + Urea | 27.5$^{hi}$ | 1.84$^a$ | 1488$^{gh}$ | 0.23$^{hij}$ | 33 | 2369$^b$ |
| | Compost + Urea | 18.8$^{hi}$ | 1.02$^d$ | 946$^h$ | 0.13$^j$ | 18 | 2756$^a$ |
| | Digestate + Urea | 25.5$^{hi}$ | 0.91$^{de}$ | 1142$^{gh}$ | 0.17$^{ij}$ | 24 | 2615$^a$ |
| | Compost + AS | 15.6$^i$ | 1.05$^{de}$ | 848$^h$ | 0.12$^j$ | 3 | 2740$^a$ |
| | Urea | 32.5$^{ghi}$ | 1.61$^b$ | 1586$^{gh}$ | 0.24$^{hij}$ | 16 | 2461$^b$ |
| | Control | 16.7$^i$ | 0.45$^{hij}$ | 706$^h$ | 0.25$^{fghi}$ | 7 | 1029$^f$ |
| Water | | ** | ** | ** | ** | | ** |
| Fertiliser | | ** | ** | ** | ** | | * |
| Water x Fertiliser | | ** | * | ** | ** | | NS |

Within the column, the values with different letters are significantly different at the p < 0.05 level. The results from two-way ANOVA indicate relationships as NS (not significant) or significant at * P < 0.05 or ** P < 0.01.
FAO, 2016. FAO rice price update http://www.fao.org/economic/est/publications/rice-publications/the-fao-rice-price-update/en/.
† March 2018 European Climate exchange CO$_2$ eq price of US$ 12.75 Mg$^{-1}$.
‡ Yearly average price of rice in 2016 in Asia US$ 365 Mg$^{-1}$ by FAO Rice Market Monitor.

## 4. Discussion

### 4.1. Factors affecting grain yield

The e-AWD water regime increased rice yield by 4–16 % overall compared to the corresponding treatments in the conventional CF water regime (Table 2). This yield gain (which was significant for two individual fertiliser application treatments) ties well with earlier studies (Ramasamy et al., 1997; Wang et al., 2000; Qin et al., 2010; Feng et al., 2013; Liu et al., 2013a,b), but contrasted with others that have reported yield reductions due to conventional AWD (Bouman and Tuong, 2001; Towprayoon et al., 2005; Linguist et al., 2015). Rice sensitivity to non-saturated soil conditions is a major concern (Bouman and Tuong, 2001), therefore it is not surprising that the yield response to AWD varies greatly in light of the large array of water management systems regarded as AWD. Both timing and severity are crucial to grain yields (Boonjung and Fukai, 1996; LaHaue et al., 2016). The present e-AWD regime included a "safe AWD threshold" for rewetting (a water level of 15 cm below the soil surface; > 10 kPa), which is promoted by the IRRI to prevent a yield loss (Bouman, 2007). Numerous field observations have supported this threshold and have also found no negative effect of AWD on yield (Dong et al., 2012; Yao et al., 2012; Pandey et al., 2014). Although mitigation of $CH_4$ emissions might be increased with increasing the severity of drainage (Wassmann et al., 2000), it can have a negative effect on grain yield (Lu et al., 2000).

Furthermore, other parameters might also influence the increase in grain yield. The wetting and drying cycles in e-AWD reinforce the air exchange between the soil and the atmosphere, thus supplying sufficient oxygen to the root system and accelerating soil organic matter mineralisation, which should result in more actively available nutrients for rice growth. The large root dry matter with high root activity found in this study (Fig. 4) implies that there is a strong water and nutrient absorption capacity, which tends to favour high grain production. Practice like AWD reported to have better resistance to lodging (bending over) of stems, attributable to better anchoring of well-developed roots. A moderate drainage regime has been shown to enhance root growth, facilitate the remobilisation of plant reserve C to the grains, accelerate grain filling and improve grain yield (Yu et al., 2010). Moreover, this study's SEM model also illustrated a positive relationship between root biomass and grain yield, which may explain the increased yield in e-AWD plots. Furthermore, a meta-analysis by Linquist et al. (2012) found that AWD resulted in greater yield benefits in soils with SOC > 1 %, as high SOC is associated with lower bulk density, higher aggregate stability, high porosity and improved structure, all of which increase water-holding capacity and plant-available water (Murphy, 2014). Finally, Kader et al. (2013) found that SOC is also positively correlated with N mineralisation in aerobic rice soil, but not in anaerobic rice soil, which suggests that N availability could increase under AWD systems conducted in high SOC soils. This is further corroborated by the present findings (Fig. 4), which indicate higher availability of $NH_4^+$ in e-AWD throughout the season.

### 4.2. Methane emissions

Methane emission is the outcome of the opposing processes of methanogenesis and methanotrophy (Le Mer and Roger, 2001), The results of the present study showed that in flooded rice systems the former is greater than the latter, generating positive $CH_4$ emissions during the rice growing period. In general, low methane effluxes were found at the beginning of the growing season, possibly due to $H_2$ competition between methanogenic, sulfate-reducing and denitrifier bacteria due to $NO_3^-$ and $SO_4^-$ availability (Le Mer and Roger, 2001). There was a gradual increase in methane emission as the plants aged and there were two peaks of $CH_4$ efflux (Fig. 1), one at the vegetative stage around 21 DAT and another at the reproductive stage, from continuously flooded plots with different organic and combination fertilisers. The initial peak of $CH_4$ efflux at the vegetative stage was probably due to the decomposition of the applied organic matter and the native easily mineralisable organic matter in the soil, a decrease in Eh due to the development of anaerobic soil conditions, and rapid growth of the rice plants that facilitated the plant-mediated transport of $CH_4$. The second peak during the reproductive period of the rice plants from continuously flooded plots was likely to be associated with the large supply of litter and root exudates and highly reduced conditions in the rice rhizosphere. This addition of carbon sources on top of the slowly mineralised organic fertilisers under the CF water regime probably increase the substrate availability for methanogens, which subsequently increase $CH_4$ production and ultimately emission to the atmosphere (Kimura et al., 2004; Gaihre et al., 2011). However, in the e-AWD treatment the initial peak of $CH_4$ was substantially lower, while the second peak was absent. During drying, soil changes from saturated to unsaturated conditions causing physical release of trapped $CH_4$ which often believe to be reason for this second spike/peak (Linquist et al., 2015). In the present study, $CH_4$ emissions were found to be significantly higher from the FYM, Digestate and Compost treatments under the conventional CF practice, which is directly in line with previous findings (Adhya et al., 2000; Zou et al., 2005; Naser et al., 2007; Ma et al., 2009).

The degree of $CH_4$ emission and reduction facilitated by e-AWD varied between the different fertiliser treatments. FYM had the highest and Compost the lowest seasonal $CH_4$ emissions in both water regimes. The FYM and biogas digestate used in this study showed high $CH_4$ emissions, probably due to large amounts of readily available C accelerating the reduction process and also supplying methane-producing substrates to methanogens (Wang et al., 2000; Pathak et al., 2003; Ma et al., 2009; Khosa et al., 2010). However, the composting process possibly stabilised the labile C from manure compost, thus reducing the potential for emissions. It has been shown that composting FYM can result a 75 % reduction in $CH_4$ emissions compared to uncomposted FYM (Chen et al., 2011). The present study was in accordance with that study because it indicated that conversion of FYM to compost reduced $CH_4$ emissions by 51 % in CF and by 61 % in the e-AWD water regime.

Overall, the e-AWD water regime reduced seasonal $CH_4$ emissions from the FYM, Compost and Digestate treatments by 69, 76 and 85 % respectively. These large reductions are consistent with previous findings (Tyagi et al., 2010; Itoh et al., 2011; Yang et al., 2012; Ly et al., 2013; Ma et al., 2013; Liu et al., 2014; Pandey et al., 2014; Linquist et al., 2015; Xu et al., 2015). Earlier studies of AWD, however, have mainly focused on the artificial fertiliser or partial organic matter supplement. The present study is the first comprehensive study in which various forms of commonly used organic fertilisers alone and in combination with synthetic mineral fertilisers have been tested side by side.

Linquist et al. (2011) reported in a meta-analysis that on average drainage reduces $CH_4$ emissions by 49.5 %. The IPCC Tier 1 system considers that a single drainage reduces $CH_4$ emissions by 40 % and multiple drainage e.g. AWD practice reduces by 48 % (Yan et al., 2005; Lasco et al., 2006), well below the results of the present e-AWD regime. The highest emission reduction due to e-AWD was found in the biogas digestate treatment, which also showed a significantly lower DOC content. DOC may be the key source of carbon for $CH_4$ production, resulting in a strong positive correlation between the $CH_4$ emissions and seasonal pattern of DOC concentrations, especially in the root zone (Lu et al., 2000). The present findings suggest that the presence of a more readily mineralisable, residue-derived DOC pool under continuous flooding was most likely linked to greater availability of substrate for methane production, particularly at the start of the rice growing season (Watanabe et al., 1999; Katoh et al., 2005). Thus, long early-season drainage (LED) (Islam et al., 2018) practised in the e-AWD water regime results in a remarkable emission reduction by stabilising DOC early in the season. The present SEM model also indicated that decreasing DOC by increasing soil redox potential through water management could lead to low $CH_4$ emissions.

Compared to organic fertiliser-only treatments, the combined mineral and organic fertiliser treatments were found to offer the most promising results in terms of decreased methane emissions. Compost + ammonium sulfate in particular showed the lowest $CH_4$ emissions in both the CF and the e-AWD water regimes. The use of fertilisers or amendments containing sulfate has been proposed as a way to mitigate $CH_4$ emissions (Linquist et al., 2012) and the present findings were in good agreement with this proposal. The way in which sulfate and other electron acceptors suppress methanogenesis can be explained by three possible mechanisms. Firstly, the electron acceptors other than $CO_2$, especially sulfate, may lead to bacterial competition that is unfavourable to methanogens while reducing concentrations of substrate to an amount that is too small for methanogenesis. Secondly, the presence of electron acceptors could increase the soil redox potential which is too high thus not favourable for methanogenesis. Thirdly, they could also be toxic for methanogens and thus decrease $CH_4$ production (Wassmann et al., 2000; Le Mer and Roger, 2001; Linquist et al., 2012).

The greater reduction in $CH_4$ emissions in the e-AWD treatments during the rice growing season in the present study can therefore be attributed to the stabilisation of labile C early in the season, which lowers the height of the first peak, followed by negligible $CH_4$ emissions later in the season, including the absence of end-of-season $CH_4$ spikes commonly found in the CF water regime.

### 4.3. Nitrous oxide emissions

The temporal $N_2O$ emission patterns and intensities found in this study were consistent with previous studies that show $N_2O$ emissions to be very minor compared to $CH_4$ emissions in flooded rice systems (Akiyama et al., 2005; Zou et al., 2007; Liu et al., 2010; Qin et al., 2010). All the organic fertilisers resulted in increased $N_2O$ emissions compared to the unfertilised control. Similar to $CH_4$ emissions, FYM and compost had the highest and lowest emissions, respectively, in the CF regime, reflecting similar trends in DOC content and suggesting that differences were due to different amounts of energy sources for denitrifiers (Murphy et al., 2003; Ju et al., 2011). However, $N_2O$ emissions from organic fertilisers were lower than the synthetic N fertiliser (Urea) control, which is line with previous studies (Ball et al., 2004; Meijide et al., 2007; Ding et al., 2013). In contrast, a number of earlier studies have also reported that manure application increases $N_2O$ emissions in farmed soils compared to chemical N fertilisers (Baggs et al., 2000; Rochette et al., 2004; Zhou et al., 2014).

Due to the introduction of aerobic cycles, alternate wetting and drying practices often lead to increased $N_2O$ emissions (Akiyama et al., 2005; Zou et al., 2007), as also observed with the e-AWD water regime in the present study (Fig. 2). Previous studies have also reported higher $N_2O$ emissions due to drainage events and from intermittent irrigations compared to the conventional CF practice (Yan et al., 2000; Nishimura et al., 2004; Towprayoon et al., 2005; Zou et al., 2005, 2009; Jiao et al., 2006; Jiang et al., 2019). The SEM model shown the positive relationship of redox potential, $NH_4^+$ and $NO_3^-$ with the increase in $N_2O$ emissions. The aerobic soil condition created by the e-AWD practice increases soil redox potential which makes the conditions favourable for $N_2O$ production. Thus, the introduction of aerobic periods could allow for nitrification of fertiliser $NH_4^+$ to $NO_3^-$ and thus greater potential for denitrification losses in the subsequent flooding phase (Buresh et al., 2008). Indeed, the e-AWD introduced very large fluctuations in $NH_4^+$ and $NO_3^-$ concentrations (Fig. 3), indicating significant intermittent ammonification and nitrification activity. In general, the organic fertilisers led to somewhat higher seasonal $N_2O$ emissions in the e-AWD water treatment compared with the conventional CF (Table 3), with the highest emission being observed from the compost treatment. The effect of synthetic N fertiliser (Urea) on $N_2O$ emission was most profound in both the CF and e-AWD water regimes. Directly in line with the previous findings, significant $N_2O$ peaks were observed in all the mineral and combination treatments immediately after synthetic N fertiliser application (Pathak et al., 2002; Zou et al., 2005; Pandey et al., 2014). After topdressing of mineral N, nitrification may have enhanced in aerobic zone due to readily available N substrate from fertilisation and subsequent denitrification in anaerobic zones of the rhizosphere, leading to accelerated emissions of $N_2O$ (Pandey et al., 2014). Application of urea and ammonium sulfate may affect $N_2O$ emissions differently due to the difference in their nitrification rates and reverse effects on soil pH (Burger and Venterea, 2011). However, we found similar $N_2O$ emissions from the combination of compost with urea or ammonium sulfate, which is consistent with what has been found in previous studies (Bouwman et al., 2002). Nevertheless, relative to the total annual emissions, the contribution of $N_2O$ emissions was < 20 %. Therefore, the overall mitigation of global warming potential was not eliminated by the increase in $N_2O$ emissions in this study (LaHue et al., 2016).

### 4.4. Global warming potentials (GWP), yield-scaled GWPs and carbon payment

In this study, $CH_4$ emissions represented 90–98 % of the area-scaled GWP across various fertiliser regimes under continuous flooding, and moderate increases in $N_2O$ emissions under e-AWD did not result in elimination of its overall mitigation of global warming potential (Tsuruta et al., 1998; Kurosawa et al., 2007; LaHue et al., 2016). This result ties well with previous studies which reported a relatively small role of $N_2O$ emissions for conventionally flooded systems (Linquist et al., 2012; Pittelkow et al., 2014b). The reduction of methane emissions by 69–85 % in the e-AWD water regime, therefore, resulted in a massive decrease in area-scaled GWP of 69–83 % for treatments receiving organic fertilisers only and by 63 % in combined organic and mineral fertiliser applications. This reduction in GWP was relatively high compared to previous studies, possibly reflecting the very effective early stabilisation of DOC through early drainage in the e-AWD water regime that was applied here.

In addition to water management practices such as the e-AWD regime, GWP also depends on the quality of the organic fertilisers. For example, pre-treated manures, such as compost and digestate, have the potential to be a more optimal organic fertiliser option because they have very low GWP compared to raw manures, i.e. FYM. When comparing $CO_2$-eq. GWP from 80 kg of FYM with an equal amount of mineral fertiliser, e.g. urea, FYM had 74 % higher $CO_2$ eq. GWP in the CF water regime, indicating its unsuitability under a conventional system. However, under the e-AWD water regime the $CO_2$ eq. GWP of FYM was found to fall from 74 % higher than the equal amount of urea to 20 % lower, indicating the strong effect of water regime in FYM. However, pre-treated organic fertilisers such as compost were shown to have only 7 % higher $CO_2$ eq. GWP compared to the same amount of urea in the CF water regime, which was 83 % lower than urea under the e-AWD water regime. Furthermore, by increasing root growth and N availability from the fertilisers applied throughout the rice-growing season, especially in the critical reproductive stage, the e-AWD water regime successfully facilitated the observed yield increase in contrast to many previous AWD studies.

For conventionally flooded treatments, the organic fertilisers increased $CH_4$ emissions but did not affect rice yield in comparison to combined synthetic and organic fertilisers, leading to yield-scaled GWP that was higher than previously reported. The e-AWD water regime significantly decreased yield-scaled GWPs owing to the reduced $CH_4$ emissions and increased yield, especially in treatments with combinations of organic and synthetic fertilisers.

It is reported that the adoption of GHG mitigation strategies is low so far, which highlights the significance of providing incentives to achieve greater adoption of such practice. To make mitigation strategies such as e-AWD more attractive in all scenarios, consideration could be given to compensating farmers for emission reductions compared to traditional baseline production as in CF. This can be done either as a carbon (C) credit payment based on the CO2-quota market (where available, e.g. March 2018 European Climate exchange CO2 eq. price of US$ 12.75 Mg-1) or as a tax on farmers who implement traditional flooding that ultimately has the same economic outcome. The e-AWD practice appears to be an economically viable option even without further economic incentives as no significant difference in yield was found between the conventional and e-AWD practice. On top of the equivalent yield, e-AWD have added benefit due to its water-saving (29 %) potential. It is noteworthy that in a C payment scheme at current CO2 and rice prices, the savings in CO2-eq. would add no more than 3 % extra revenue (max $ 78 out of $ 2518 rice value), while the yield increase from e-AWD for the organic fertilisers only or with synthetic fertilisers would result in extra revenue of 5–16 % ($ 110–365 out of $ 2518). However, if C credit payments were considered, the present e-AWD water regime would be the most economically attractive option by increasing profits with both the organically fertilised and combination fertiliser treatments (Table 3). To implement the e-AWD practice, farmers will mainly require an AWD field water tube for water level monitoring which can be easily made with locally available materials like bamboo or plastic pipes. As e-AWD is a low tech easily adaptable practice with low economic investment, its adoption could be sustainable for farmers in the long term.

### 4.5. Heavy metals in rice grain (arsenic, lead and cadmium)

The arsenic content of rice and associated potential health concerns have recently been highlighted, especially in south-east Asian countries where rice is a daily staple food (Williams et al., 2007; Zhu et al., 2008; Banerjee et al., 2013; Linquist et al., 2015). In present study, grain As concentrations for the treatments receiving organic fertilisers under continuous flooding averaged about 410 µg kg$^{-1}$ (Table 2), which is higher than the average for that variety reported in recent studies in California (Linquist et al., 2015; LaHau et al., 2016), but lower than studies from Bangladesh and China (Williams et al., 2007; Norton et al., 2012). However, these studies were from systems receiving inorganic fertiliser. No studies receiving organic fertiliser were found in the literature. The WHO and FAO of the United Nations have set a voluntary recommended As limit for polished (milled) rice at 200 µg kg$^{-1}$ (Codex Alimentarius Commission, 2014). Therefore, it is essential to have agronomic strategies in rice production systems receiving organic fertiliser that can reduce uptake of As in rice grain. The e-AWD water regime resulted in a 57 % reduction in As concentration on average, from 410 µg kg$^{-1}$ to 177 µg kg$^{-1}$, i.e. below the recommended limit. This reduction is in line with previous studies, for example, Somenahally et al. (2011) found a 41 % reduction on average in total grain As from paddy rice grown in Texas on fields that were flooded intermittently, while a reduction of 40 % on average was reported in rice fields in California (Linquist et al., 2015). However, this is the first study to confirm such a high As reduction potential from the e-AWD water treatment in organically fertilised rice systems.

It was hypothesised that this effect on As is related to redox chemistry. The continuously flooded plots have more sulfate, Fe(III)- and As(V)-reducing bacteria than the non-flooded treatments such as the e-AWD water regimes (Das et al., 2016). The role of these bacteria in the completely flooded system is to enhance the As(III) concentration in soil pore water (Das et al., 2013), which is the dominant species taken up by rice roots (Chen et.al, 2005). Reduced conditions combined with a relatively high dissolved organic carbon content in the rhizosphere soil of flooded paddies, owing to the addition of organic manure/residue, are likely to favour the growth and activity of these reducing bacteria, As release to the soil solution, and subsequent As(III) uptake by the plant (Norton et al., 2013; Jia et al., 2001). Aerobic periods introduced by e-AWD would favour the oxidation of As(III), leading to formation of As(V), which is less toxic and less bioavailable and has a much higher affinity for Fe-(hydr)oxides than As(III) (Takahashi et al., 2004; LaHue et al., 2016).

The bioavailability and mobility of Cadmium depend on soil organic matter content and pH. The soil pH can decrease and release metals due to oxidation of reduced soil components following soil drainage or re-drying episodes (Fulda et al., 2013). However, the application of organic matter like in organic production systems can retard the oxidation and buffer the pH change (Yuan et al., 2016). In this regard, Yuan et al. (2019) reported that application of organic matter is the most effective option for immobilising dissolved soil cadmium in anaerobic soil conditions like found in rice production. Such application of organic matter provides electron donors to iron-reducing bacteria which can restrict Cd remobilization following soil drainage (Yuan et al., 2019). In this study, we found that the e-AWD water regime reduced grain Cd concentration in compost and digestate treatment which is consistent with previous studies by Honma et al. who also found intermittent irrigation simultaneously reduce As and Cd uptake in rice grain. It has been reported that the moderate alternate drying enhances the diversity index of the rhizosphere bacterial community structure which may reduce the uptake of heavy metals like Cd in rice grain whereas severe alternate drying and wetting showed the opposite effect (Peralta et al., 2014; Zhang et al., 2019). Thus, the application of organic matter and moderate drying threshold of e-AWD might have played a major role in the lower accumulation of Cd in grain. Under moderate drying conditions in e-AWD, cadmium ions ($Cd^{2+}$) may precipitate as cadmium sulfate. This may result in a reduction of the soil solution concentration of cadmium (Barrett and McBride, 2007) and in the lower Cd concentration in rice grain in the present study. However, we found no significant difference in Cd concentration between the two water regimes for combination treatments. On the other hand, a significant difference was found in Pb concentration between the two water regimes in both organic and combination treatments. On average, the e-AWD water regime resulted in a 72 % and 56 % reduction in grain lead concentrations in the organic and combination treatments, respectively. The common safety threshold of Pb and Cd is set at 200 µg kg$^{-1}$ for rice grain (SCOOP, 2004; Chinese Food Standards Agency, 2005; FAO/WHO, 2011). Although the total grain Pb and Cd concentrations in the present study were always below the safety threshold limit, e-AWD might potentially help to reduce the Pb content in areas that have problems with this metal, as previous studies have shown that uptake of these metals in rice grain in the polluted areas can dangerously exceed the recommended safety limit, rising to as much as 1.00 mg kg$^{-1}$ (Liu et al., 2003; Cheng et al., 2006; Fangmin et al., 2006; Yang et al., 2006; Fu et al., 2008; Williams et al., 2009, 2012; Liu et al., 2013a,b).

## 5. Management implications and future directions

This study found that the environmental impacts of rice cultivation can be minimised by e-AWD water regime without compromising optimal agronomic performance. The AWD irrigation regime has been shown in numerous previous studies to have the potential to reduce both water use and GHG emissions, but this has frequently been at the loss of rice grain yield. The present proposed e-AWD irrigation system appears to offer the optimum balance with respect to both yield and GHG emissions, with the potential of reducing heavy metal concentrations compared to continuous flooding. A key pathway through which e-AWD decreases $CH_4$ emissions is early-season stabilisation of DOC during early drainage, removing substrate for subsequent methanogenesis.

These findings further suggest that rice production under conventional flooding with the application of mainly organic fertilisers (as is done in certified organic farming for example) may not be an environmentally sustainable option, adding high GHG emissions to the perils of considerable water use and a possibly increased uptake of heavy metals. Therefore, it is suggested that rice producers comply with organic certification schemes, always try to implement e-AWD-like water regimes, and use pre-treated manure such as compost and digestate as fertilisers instead of raw manure such as farmyard manure. However, in many regions larger-scale certified organic rice production is limited by the availability of sufficient quantities of organic manure, lower yields and increased labour requirements. In the absence of sufficient quantities of organic fertiliser, combination treatments of organic and synthetic fertilisers (as for example the 50:50 combination used in the present study) could be a good option, also ensuring an NPKS application that is more in balance with crop demand. Ammonium sulfate combined with compost in particular showed very encouraging results. However, it is recognised that these combinations would not be an option for certified organic production. Finally, it should be emphasised that implementing any form of alternate wetting and drying practice in the wet season in the tropical region may be very challenging (Adhya et al., 2014). This implies that the use of composted manure is preferable to farmyard manure or digestate during the wet season. This experiment was conducted in the dry season when soil moisture condition can be effectively controlled. The e-AWD practice which has shown multiple environmental and agronomic benefit is a good proof of concept, however, it has to be tested over multiple years as well as under farmers field conditions. We assume that in actual farmers field, the mitigation of GHG emissions and reduction of grain heavy metals could be little lower as some farmers may face important technical and practical constraints to implementing such improvements such as not being able to manage their water reliably, lower soil fertility and carbon stock of their field, uneven land with pockets with excessive wetting or drying, continuous heavy rainfall in wet season etc. Therefore, future research should focus on adapting e-AWD to field scales. This study demonstrates that adoption e-AWD practice in organic production systems has the promise to achieve numerous environmental as well as agronomic goals and thus should be considered as a feasible option for simultaneous reduction of GHG, water use and rice grain heavy metal uptake (e.g. As, Pb, Cd) without affecting yield, achieving environmental safety benchmarks in addition to existing health safety benefits.

**Declaration of Competing Interest**

The authors declare that they have no known competing financial interests or personal relationships that could have appeared to influence the work reported in this paper.

**Acknowledgements**

This research work was funded by the European Commission (agreement no. 2013-008) under the Agricultural Transformation by Innovation, Erasmus Mundus joint Doctoral fellowship programme. This work was further supported by the Climate and Clean Air Coalition (CCAC) and the CGIAR Research Program on Climate Change, Agriculture and Food Security (CCAFS) which is carried out with support from CGIAR Trust Fund Donors and through bilateral funding agreements. For details please visit https://ccafs.cgiar.org/donors. The views expressed in this document cannot be taken to reflect the official opinions of these organizations. This work also received support through the CGIAR RICE agrifood research programme. We express our appreciation to all field staff of IRRIs Climate Change group and the Crop and Environmental Sciences Division (CESD) for the experimental setup and managing field operations, and laboratory staff for the assistance in GC analysis of gas samples. We also acknowledge the assistance of all staff at the IRRI Zeigler Experiment Station. Finally, we would also like to acknowledge the important contribution made by all the laboratory staff at the Copenhagen Plant Science and Soil group laboratory, the Soil Biology group and the CBLB laboratory of Wageningen University in The Netherlands.

**References**

Abalos, D., Jeffery, S., Drury, C.F., Wagner-Riddle, C., 2016. Improving fertilizer management in the US and Canada for N2O mitigation: understanding potential positive and negative side-effects on corn yields. Agric. Ecosyst. Environ. 221, 214–221.

Adhya, T., Bharati, K., Mohanty, S., Ramakrishnan, B., Rao, V., Sethunathan, N., Wassmann, R., 2000. Methane emission from rice fields at Cuttack, India. Nutr. Cycling Agroecosyst. 58, 95–105.

Adhya, T.K., Linquist, B., Searchinger, T., Wassmann, R., Yan, X., 2014. Wetting and drying: reducing greenhouse gas emissions and saving water from rice production. Installment 8 of Creating a Sustainable Food Future. pp. 1–28.

Ahn, J., Choi, M., Kim, B., Lee, J., Song, J., Kim, G., Weon, H., 2014. Effects of water-saving irrigation on emissions of greenhouse gases and prokaryotic communities in rice paddy soil. Microb. Ecol. 68, 271–283.

Akiyama, Hiroko, Yagi, Kazuyuki, Yan, Xiaoyuan, 2005. Direct N2O emissions from rice paddy fields: Summary of available data. Global Biogeochem. Cycles 19.

Baggs, E.M., Rees, R.M., Smith, K.A., Vinten, A.J.A., 2000. Nitrous oxide emission from soils after incorporating crop residues. Soil Use Manage. 16, 82–87.

Ball, B.C., McTaggart, I.P., Scott, A., 2004. Mitigation of greenhouse gas emissions from soil under silage production by use of organic manures or slow-release fertilizer. Soil Use Manage 20, 287–295.

Banerjee, M., Banerjee, N., Bhattacharjee, P., Mondal, D., Lythgoe, P.R., MartÃnez, M., Pan, J., Polya, D.A., Giri, A.K., 2013. High arsenic in rice is associated with elevated genotoxic effects in humans. Sci. Rep. 3, 2195.

Belder, P., Bouman, B., Cabangon, R., Lu, G., Quilang, E., Li, Y., Spiertz, J., Tuong, T., 2004. Effect of water-saving irrigation on rice yield and water use in typical lowland conditions in Asia. Agric. Water Manage. 65, 193–210.

Boonjung, H., Fukai, S., 1996. Effects of soil water deficit at different growth stages on rice growth and yield under upland conditions. 2. Phenology, biomass production and yield. Field Crops Res. 48, 47–55.

Bouman, B.A.M., 2007. Water management in irrigated rice: coping with water scarcity. Int. Rice Res. Inst. Anonymous, Philippines.

Bouman, B., Tuong, T.P., 2001. Field water management to save water and increase its productivity in irrigated lowland rice. Agric. Water Manage. 49, 11–30.

Bouwman, A., Boumans, L., Batjes, N., 2002. Modeling global annual N2O and NO emissions from fertilized fields. Global Biogeochem. Cycles 16, 1080.

Buresh, R., Reddy, K., van Kessel, C., 2008. Nitrogen transformations in submerged soils. Nitrogen in Agricultural Systems, Agronomy Monograph 49. American Society of Agronomy. Crop Science Society of America, Soil Science Society of America, Madison.

Burger, M., Venterea, R.T., 2011. Effects of nitrogen fertilizer types on nitrous oxide emissions. Anonymous Understanding Greenhouse Gas Emissions From Agricultural Management. American Chemical Society.

Carrijo, D.R., Lundy, M.E., Linquist, B.A., 2017. Rice yields and water use under alternate wetting and drying irrigation: a meta-analysis. Field Crops Res. 203, 173–180.

Chen, Z., Zhu, Y., Liu, W., Meharg, A., 2005. Direct evidence showing the effect of root surface iron plaque on arsenite and arsenate uptake into rice (Oryza sativa) roots. New Phytol. 165, 91–97.

Cheng, F., Zhao, N., Xu, H., Li, Y., Zhang, W., Zhu, Z., Chen, M., 2006. Cadmium and lead contamination in japonica rice grains and its variation among the different locations in southeast China. Sci. Total Environ. 359, 156–166.

Chinese Food Standards Agency, 2005. Maximum Levels of Contaminants in Food, GB 2762-2005. Chinese Food Standards Agency, China.

D'ilio, S., Alessandrelli, M., Cresti, R., Forte, G., Caroli, S., 2002. Arsenic content of various types of rice as determined by plasma-based techniques. Microchem. J. 73, 195–201.

Das, S., Jean, J., Kar, S., Chakraborty, S., 2013. Effect of arsenic contamination on bacterial and fungal biomass and enzyme activities in tropical arsenic-contaminated soils. Biol. Fertility Soils 49, 757–765.

Das, S., Chou, M., Jean, J., Liu, C., Yang, H., 2016. Water management impacts on arsenic behavior and rhizosphere bacterial communities and activities in a rice agro-ecosystem. Sci. Total Environ. 542, 642–652.

de Vries, F.T., Bardgett, R.D., 2016. Plant community controls on short-term ecosystem nitrogen retention. New Phytol. 210, 861–874.

Diacono, M., Montemurro, F., 2010. Long-term effects of organic amendments on soil fertility. A review. Agron. Sustain. Dev. 30, 401–422.

Ding, W., Luo, J., Li, J., Yu, H., Fan, J., Liu, D., 2013. Effect of long-term compost and inorganic fertilizer application on background N2O and fertilizer-induced N2O emissions from an intensively cultivated soil. Sci. Total Environ. 465, 115–124.

Dong, N.M., Brandt, K.K., Sørensen, J., Hung, N.N., Hach, C.V., Tan, P.S., Dalsgaard, T., 2012. Effects of alternating wetting and drying versus continuous flooding on fertilizer nitrogen fate in rice fields in the Mekong Delta. Vietnam. Soil Biology and Biochemistry 47, 166–174.

Fangmin, C., Ningchun, Z., Haiming, X., Yi, L., Wenfang, Z., Zhiwei, Z., Mingxue, C., 2006. Cadmium and lead contamination in japonica rice grains and its variation among the different locations in southeast China. Sci. Total Environ. 359, 156–166.

FAO, 2018. AQUASTAT Data. (accessed 04.03.2018).. http://www.fao.org/nr/water/aquastat/data/query/index.html?lang=en.

FAO/WHO, 2011. Food Standards Programme Codex Committee on Contaminants in Foods. Fifth Session. FAO/WHO, The Hague, The Netherlands, pp. 21–25.

FAOSTAT, 2014. Food and Agriculture Organization of the United Nations, Statistics Division. Area Harvested (hectares), Rice, World Total, 2012. Retrieved from the FAOSTAT database. http://faostat3.fao.org (accessed 09.04.2017). FAOSTAT.

Feng, J., Chen, C., Zhang, Y., Song, Z., Deng, A., Zheng, C., Zhang, W., 2013. Impacts of cropping practices on yield-scaled greenhouse gas emissions from rice fields in China:

a meta-analysis. Agric. Ecosyst. Environ. 164, 220–228.

Foley, J.A., Ramankutty, N., Brauman, K.A., Cassidy, E.S., Gerber, J.S., Johnston, M., Mueller, N.D., O'Connell, C., Ray, D.K., West, P.C., 2011. Solutions for a cultivated planet. Nature 478, 337.

Fu, J., Zhou, Q., Liu, J., Liu, W., Wang, T., Zhang, Q., Jiang, G., 2008. High levels of heavy metals in rice (Oryzasativa L.) from a typical E-waste recycling area in southeast China and its potential risk to human health. Chemosphere 71, 1269–1275.

Fulda, B., Voegelin, A., Kretzschmar, R., 2013. Redox-controlled changes in cadmium solubility and solid-phase speciation in a paddy soil as affected by reducible sulfate and copper. Environ. Sci. Technol. 47, 12775–12783.

Gaihre, Y., Tirol-Padre, A., Wassmann, R., Aquino, E., Pangga, G., Sta-Cruz, P., 2011. Spatial and temporal variations in methane fluxes from irrigated lowland rice fields. Philipp. Agric. Sci. 94, 335–342.

Godfray, H.C., Pretty, J., Thomas, S.M., Warham, E.J., Beddington, J.R., 2011. Global food supply. Linking policy on climate and food. Science 331, 1013–1014.

Gómez-Muñoz, B., Magid, J., Jensen, L.S., 2017. Nitrogen turnover, crop use efficiency and soil fertility in a long-term field experiment amended with different qualities of urban and agricultural waste. Agric. Ecosyst. Environ. 240, 300–313.

Grace, J.B., 2006. Structural Equation Modeling and Natural Systems. Cambridge University Press, Cambridge.

Henry, A., Gowda, V.R.P., Torres, R.O., McNally, K.L., Serraj, R., 2011. Variation in root system architecture and drought response in rice (Oryza Sativa): Phenotyping of the Oryzasnp panel in rainfed lowland fields. Field Crops Res.

Hoekstra, A.Y., Mekonnen, M.M., 2012. The water footprint of humanity. Proc. Natl. Acad. Sci. U.S.A. 109, 3232.

Horwitz, W., 1982. Evaluation of analytical methods used for regulation of foods and drugs. Anal. Chem. 54, 67–76.

Hou, H., Peng, S., Xu, J., Yang, S., Mao, Z., 2012. Seasonal variations of $CH_4$ and $N_2O$ emissions in response to water management of paddy fields located in Southeast China. Chemosphere 89, 884–892.

Houba, V.J.G., Temminghoff, E.J.M., Gaikhorst, G.A., van Vark, W., 2000. Soil analysis procedures using 0.01 M calcium chloride as extraction reagent. Commun. Soil Sci. Plant Anal. 31, 9–10. https://doi.org/10.1080/00103620009370514. 1299-1396.

Inubushi, K., Brookes, P.C., Jenkinson, D.S., 1991. Soil microbial biomass C, N and ninhydrin-N in aerobic and anaerobic soils measured by the fumigation-extraction method. Soil Biol. Biochem. 23, 737–741.

IPCC, 2007. Intergovernmental panel on climate change. In: Parry, M.L., Canziani, O.F., Palutikof, J.P., Van der Linden, P.J., Hanson, C.E. (Eds.), Climate Change 2007: The Scientific Basis, Impacts, Adaptation and Vulnerability. Cambridge University Press, Cambridge, UK and New York, NY, USA.

IRRI, 2013. Rice Facts. International Rice Research Institute, Manila, Philippines. http://irri.org.

Islam, S.F., van Groenigen, J.W., Jensen, L.S., Sander, B.O., de Neergaard, A., 2018. The effective mitigation of greenhouse gas emissions from rice paddies without compromising yield by early-season drainage. Sci. Total Environ. 612, 1329–1339.

Itoh, M., Sudo, S., Mori, S., Saito, H., Yoshida, T., Shiratori, Y., Suga, S., Yoshikawa, N., Suzue, Y., Mizukami, H., Mochida, T., Yagi, K., 2011. Mitigation of methane emissions from paddy fields by prolonging midseason drainage. Agric. Ecosyst. Environ. 141, 359–372.

Liu, J., Ma, X., Wang, M., Sun, X., 2013a. Genotypic differences among rice cultivars in lead accumulation and translocation and the relation with grain Pb levels. Ecotoxicol. Environ. Saf. 90, 35–40.

Liu, L., Chen, T., Wang, Z., Zhang, H., Yang, J., Zhang, J., 2013b. Combination of site-specific nitrogen management and alternate wetting and drying irrigation increases grain yield and nitrogen and water use efficiency in super rice. Field Crops Res. 154, 226–235.

Jia, Z., Cai, Z., Xu, H., Li, X., 2001. Effect of rice plants on $CH_4$ production, transport, oxidation and emission in rice paddy soil. Plant Soil 230, 211–221.

Jiang, Y., Carrijo, D., Huang, S., Chen, J., Balaine, N., Zhang, W., van Groenigen, K.J., Linquist, B., 2019. Water management to mitigate the global warming potential of rice systems: a global meta-analysis. Field Crops Res. 234, 47–54.

Jiao, Z., Hou, A., Shi, Y., Huang, G., Wang, Y., Chen, X., 2006. Water management influencing methane and nitrous oxide emissions from rice field in relation to soil redox and microbial community. Commun. Soil Sci. Plant Anal. 37, 1889–1903.

Ju, X., Lu, X., Gao, Z., Chen, X., Su, F., Kogge, M., Römheld, V., Christie, P., Zhang, F., 2011. Processes and factors controlling $N_2O$ production in an intensively managed low carbon calcareous soil under sub-humid monsoon conditions. Environ. Pollut. 159, 1007–1016.

Kader, M.A., Sleutel, S., Begum, S.A., Moslehuddin, A.Z.M., De, n.S., 2013. Nitrogen mineralization in sub-tropical paddy soils in relation to soil mineralogy, management, pH, carbon, nitrogen and iron contents. Eur. J. Soil Sci. 64, 47–57.

Katoh, M., Murase, J., Sugimoto, A., Kimura, M., 2005. Effect of rice straw amendment on dissolved organic and inorganic carbon and cationic nutrients in percolating water from a flooded paddy soil: a microcosm experiment using 13C-enriched rice straw. Org. Geochem. 36, 803–811.

Khosa, M.K., Sidhu, B., Benbi, D., 2010. Effect of organic materials and rice cultivars on methane emission from rice field. J. Environ. Biol. 31, 281–285.

Kimura, M., Murase, J., Lu, Y., 2004. Carbon cycling in rice field ecosystems in the context of input, decomposition and translocation of organic materials and the fates of their end products ($CO_2$ and $CH_4$). Soil Biol. Biochem. 36, 1399–1416.

Kline, R.B., 2011. Principles and Practice of Structural Equation Modeling, third ed. The Guilford Press, New York, pp. 427.

Kurosawa, M., Yamashiki, Y., Tezuka, T., 2007. Estimation of energy consumption and environmental impact reduction by introducing sustainable agriculture to rice paddy field cultivation. J. Life Cycle Assess Jpn. 3, 232–238.

LaHue, G.T., Chaney, R.L., Adviento-Borbe, M.A.A., Linquist, B.A., 2016. Alternate wetting and drying in high yielding direct-seeded rice systems accomplishes multiple environmental and agronomic objectives. Agric. Ecosyst. Environ. 229, 30–39.

Lasco, R.D., Ogle, S., Raison, J., 2006. Chapter 5: cropland. In: Eggleston, S., Buendia, L., Miwa, K., Ngara, T., Tanabe, K. (Eds.), 2006 IPCC Guidelines for National Greenhouse Gas Inventories. IGES, IPCC National Greenhouse Gas Inventories Program, Japan pp. 5.1–5.66.

Le Mer, J., Roger, P., 2001. Production, oxidation, emission and consumption of methane by soils: a review. Eur. J. Soil Biol. 37, 25–50.

Linquist, B.A., Adviento-Borbe, M.A., Pittelkow, C.M., van Kessel, C., van Groenigen, K.J., 2012. Fertilizer management practices and greenhouse gas emissions from rice systems: a quantitative review and analysis. Field Crops Res. 135, 10–21.

Linquist, B.A., Anders, M.M., Adviento-Borbe, M.A.A., Chaney, R.L., Nalley, L.L., Da Rosa, E.F., Van Kessel, C., 2015. Reducing greenhouse gas emissions, water use, and grain arsenic levels in rice systems. Glob. Change Biol. Bioenergy 21, 407–417.

Liu, J., Li, K., Xu, J., Zhang, Z., Ma, T., Lu, X., Yang, J., Zhu, Q., 2003. Lead toxicity, uptake, and translocation in different rice cultivars. Plant Sci. 165, 793–802.

Liu, S., Qin, Y., Zou, J., Liu, Q., 2010. Effects of water regime during rice-growing season on annual direct $N_2O$ emission in a paddy rice–winter wheat rotation system in southeast China. Sci. Total Environ. 408, 906–913.

Lu, J., Ookawa, T., Hirasawa, T., 2000. The effects of irrigation regimes on the water use, dry matter production and physiological responses of paddy rice. Plant Soil 223, 209–218.

Ly, P., Jensen, L.S., Bruun, T.B., de Neergaard, A., 2013. Methane ($CH_4$) and nitrous oxide ($N_2O$) emissions from the system of rice intensification (SRI) under a rain-fed lowland rice ecosystem in Cambodia. Nutr. Cycling Agroecosyst. 97, 13–27.

Ma, J.F., Yamaji, N., Mitani, N., Xu, X., Su, Y., McGrath, S.P., Zhao, F., 2008. Transporters of arsenite in rice and their role in arsenic accumulation in rice grain. Proc. Natl. Acad. Sci. U.S.A. 105, 9931.

Ma, J., Ma, E., Xu, H., Yagi, K., Cai, Z., 2009. Wheat straw management affects $CH_4$ and $N_2O$ emissions from rice fields. Soil Biol. Biochem. 41, 1022–1028.

Mandal, B., Suzuki, K., 2002. Arsenic round the world: a review. Talanta 58, 201–235.

Meijide, A., Díez, J.A., Sánchez-Martín, L., López-Fernández, S., Vallejo, A., 2007. Nitrogen oxide emissions from an irrigated maize crop amended with treated pig slurries and composts in a Mediterranean climate. Agric. Ecosyst. Environ. 121, 383–394.

Mueller, N.D., Gerber, J.S., Johnston, M., Ray, D.K., Ramankutty, N., Foley, J.A., 2013. Closing yield gaps through nutrient and water management. Nature 494 (vol 490, pg 254, 2012) 390-390.

Murphy, B., 2014. Soil Organic Matter and Soil Function–Review of the Literature and Underlying Data. Department of the Environment, Canberra, Australia.

Murphy, D., Recous, S., Stockdale, E., Fillery, I., Jensen, L., Hatch, D., Goulding, K., 2003. Gross nitrogen fluxes in soil: theory, measurement and application of ˆ1ˆ5n pool dilution techniques. Adv. Agron. 79, e118.

Nalley, L., Linquist, B., Kovacs, K., Anders, M., 2015. The economic viability of alternative wetting and drying irrigation in Arkansas rice production. Agron. J. 107, 579–587.

Naser, H.M., Nagata, O., Tamura, S., Hatano, R., 2007. Methane emissions from five paddy fields with different amounts of rice straw application in central Hokkaido. Japan. Soil Sci. Plant Nutr. 53, 95–101.

NEN-EN-1484, 1997. Water: Leidraad Voor De Bepaling Van Het Gehalte Aan Totaal Organische Koolstof (TOC) En Opgelost Organisch Koolstof (DOC). Accessible at:. NEN, The Netherlands. https://www.nen.nl/NEN-Shop/Norm/NENEN-14841997-nl.htm.

Nishimura, S., Sawamoto, T., Akiyama, H., Sudo, S., Yagi, K, 2004. Methane and nitrous oxide emissions from a paddy field with Japanese conventional water management and fertilizer application. Global Biogeochem. Cycles 18.

Norton, G.J., Pinson, S.R., Alexander, J., McKay, S., Hansen, H., Duan, G.L., Rafiqul Islam, M., Islam, S., Stroud, J.L., Zhao, F.J., McGrath, S.P., Zhu, Y.G., Lahner, B., Yakubova, E., Guerinot, M.L., Tarpley, L., Eizenga, G.C., Salt, D.E., Meharg, A.A., Price, A.H., 2012. Variation in grain arsenic assessed in a diverse panel of rice (*Oryza sativa*) grown in multiple sites. New Phytol. 193, 650–664.

Norton, G.J., Adomako, E.E., Deacon, C.M., Carey, A., Price, A.H., Meharg, A.A., 2013. Effect of organic matter amendment, arsenic amendment and water management regime on rice grain arsenic species. Environ. Pollut. 177, 38–47.

Pan, Y., Bonten, L.T.C., Koopmans, G.F., Song, J., Luo, Y., Temminghoff, E.J.M., Comans, R.N.J., 2016. Solubility of trace metals in two contaminated paddy soils exposed to alternating flooding and drainage. Geoderma 261, 59–69.

Pandey, A., Mai, V.T., Vu, D.Q., Bui, T.P.L., Mai, T.L.A., Jensen, L.S., de Neergaard, A., 2014. Organic matter and water management strategies to reduce methane and nitrous oxide emissions from rice paddies in Vietnam. Agric. Ecosyst. Environ. 196, 137–146.

Paradelo, R., Villada, A., Devesa-Rey, R., Moldes, A.B., Domínguez, M., Patiño, J., Barral, M.T., 2011a. Distribution and availability of trace elements in municipal solid waste composts. J. Environ. Monit. 13, 201–211.

Paradelo, R., Villada, A., Devesa-Rey, R., Belen Moldes, A., Dominguez, M., Patino, J., Teresa Barral, M., 2011b. Distribution and availability of trace elements in municipal solid waste composts. J. Environ. Monit. 13, 201–211.

Pathak, H., Bhatia, A., Prasad, S., Singh, S., Kumar, S., Jain, M.C., Kumar, U., 2002. Emission of nitrous oxide from rice-wheat systems of Indo-Gangetic Plains of India. Environ. Monit. Assess. 77, 163–178.

Pathak, H., Prasad, S., Bhatia, A., Singh, S., Kumar, S., Singh, J., Jain, M.C., 2003. Methane emission from rice–wheat cropping system in the Indo-Gangetic plain in relation to irrigation, farmyard manure and dicyandiamide application. Agric. Ecosyst. Environ. 97, 309–316.

PCRA (Protection cathodique et revêtements associés), 2007. Recommendations for the verification of reference electrodes. Recommendations PCRA 005, Rev 1. Committee for Cathodic Protection and Associated Coatings, Centre Français De l'Anticorrosion.

Accessible at: http://www.cefracor.org/html/publications.htm.
Peralta, A.L., Ludmer, S., Matthews, J.W., Kent, A.D., 2014. Bacterial community response to changes in soil redox potential along a moisture gradient in restored wetlands. Ecol. Eng. 73, 246–253.
Petersen, S.O., Henriksen, K., Mortensen, G.K., Krogh, P.H., Brandt, K.K., Sørensen, J., Madsen, T., Petersen, J., Grøn, C., 2003. Recycling of sewage sludge and household compost to arable land: fate and effects of organic contaminants, and impact on soil fertility. Soil Tillage Res. 72, 139–152.
Pittelkow, C.M., Adviento-Borbe, M.A., Chris, K., Hill, J.E., Linquist, B.A., 2014. Optimizing rice yields while minimizing yield-scaled global warming potential. Glob. Change Biol. Bioenergy 20, 1382–1393.
Qin, Y., Liu, S., Guo, Y., Liu, Q., Zou, J., 2010. Methane and nitrous oxide emissions from organic and conventional rice cropping systems in Southeast China. Biol. Fertility Soils 46, 825–834.
Ramasamy, S., ten Berge, H.F.M., Purushothaman, S., 1997. Yield formation in rice in response to drainage and nitrogen application. Field Crops Res. 51, 65–82.
Richards, M., Sander, B.O., 2014. Alternate wetting and drying in irrigated rice: implementation guidance for policymakers and investors. Climate-Smart Agriculture Practice Brief. CGIAR Research Program on Climate Change, Agriculture and Food Security (CCAFS), Copenhagen, Denmark.
Robertson, G.P., Vitousek, P.M., 2009. Nitrogen in agriculture: balancing the cost of an essential resource. Annu. Rev. Environ. Resour. 34, 97–125.
Rochette, P., Angers, D.A., Chantigny, M.H., Bertrand, N., Côté, D., 2004. Carbon Dioxide and Nitrous Oxide Emissions following Fall and Spring Applications of Pig Slurry to an Agricultural Field. pp. 1410–1420.
Rosseel, Y., 2012. Lavaan: an r package for structural equation modeling. J. Stat. Softw. 48.
Sander, B.O., Samson, M., Buresh, R.J., 2014. Methane and nitrous oxide emissions from flooded rice fields as affected by water and straw management between rice crops. Geoderma 235, 355–362.
Schlesinger, W.H., 2009. On the fate of anthropogenic nitrogen. Proc. Natl. Acad. Sci. U.S.A. 106, 203.
SCOOP, 2004. Reports on tasks for scientific cooperation (SCOOP) task 3.2.11. Assessment of the Dietary Exposure to Arsenic, Cadmium, Lead and Mercury of the Population of the EU Member States Directorate General Health and Consumer Protection. EC European Commission.
Seck, P.A., Diagne, A., Mohanty, S., Wopereis, M.C.S., 2012. Crops that feed the world 7: rice. Food Secur. 4, 7–24.
Smith, P., Martino, D., Cai, Z., 2007. Agriculture. In: Metz, B., Davidson, O.R., Bosch, P.R., Dave, R., Meyer, L.A. (Eds.), Climate Change 2007: Mitigation. Contribution of Working Group III to the Fourth Assessment Report of the Intergovernmental Panel on Climate Change. Cambridge University Press, Cambridge, UK and New York, NY, USA, pp. 497–540.
Snyder, C.S., Bruulsema, T.W., Jensen, T.L., Fixen, P.E., 2009. Review of greenhouse gas emissions from crop production systems and fertilizer management effects. Agric. Ecosyst. Environ. 133, 247–266.
Somenahally, A.C., Hollister, E.B., Yan, W., Gentry, T.J., Loeppert, R.H., 2011. Water management impacts on arsenic speciation and iron-reducing Bacteria in contrasting rice-rhizosphere compartments. Environ. Sci. Technol. 45, 8328–8335.
Spanu, A., Daga, L., Orlandoni, A.M., Sanna, G., 2012. The role of irrigation techniques in arsenic bioaccumulation in rice (Oryza sativa L.). Environ. Sci. Technol. 46, 8333–8340.
Straathof, A.L., Chincarini, R., Comans, R.N.J., Hoffland, E., 2014. Dynamics of soil dissolved organic carbon pools reveal both hydrophobic and hydrophilic compounds sustain microbial respiration. Soil Biol. Biochem. 79, 109–116.
Takahashi, Y., Minamikawa, R., Hattori, K., Kurishima, K., Kihou, N., Yuita, K., 2004. Arsenic behavior in paddy fields during the cycle of flooded and non-flooded periods. Environ. Sci. Technol. 38, 1038–1044.
Tariq, A., Quynh Duong, Vu, Jensen, L.S., de Tourdonnet, S., Sander, B.O., Wassmann, R., Trinh Van, Mai, de Neergaard, A., 2017. Mitigating CH4 and N2O emissions from intensive rice production systems in northern Vietnam: Efficiency of drainage patterns in combination with rice residue incorporation. Agric. Ecosyst. Environ. 249, 101–111.
Tirado, R., Gopikrishna, S.R., Krishnan, R., Smith, P., 2010. Greenhouse gas emissions and mitigation potential from fertilizer manufacture and application in India. Int. J. Agric. Sustain. 8, 176–185.
Towprayoon, S., Smakgahn, K., Poonkaew, S., 2005. Mitigation of methane and nitrous oxide emissions from drained irrigated rice fields. Chemosphere 59, 1547–1556.
Tuong, T., Bouman, B., Mortimer, M., 2005. More rice, less water - Integrated approaches for increasing water productivity in irrigated rice-based systems in Asia. Plant Prod. Sci. 8, 231–241.
Tyagi, L., Kumari, B., Singh, S.N., 2010. Water management — a tool for methane mitigation from irrigated paddy fields. Sci. Total Environ. 408, 1085–1090.
USEPA, 2006. Global Mitigation of Non-CO2 Greenhouse Gases. U.S. Environmental Protection Agency, Office of Atmospheric Programs (6207J), Washington, DC.
van Groenigen, K.J., van Kessel, C., Hungate, B.A., 2013. Increased greenhouse-gas intensity of rice production under future atmospheric conditions. Nat. Clim. Chang. 3, 288–291.
Van Groenigen, J.W., Velthof, G.L., Oenema, O., Van Groenigen, K.J., Van Kessel, C., 2010. Towards an agronomic assessment of N2O emissions: a case study for arable crops. Eur. J. Soil Sci. 61, 903–913.
Vance, E.D., Brookes, P.C., Jenkinson, D.S., 1987. An extraction method for measuring soil microbial biomass C. Soil Biol. Biochem. 19, 703–707.
Vu, Q.D., de Neergaard, A., Tran, T.D., Hoang, Q.Q., Ly, P., Tran, T.M., Jensen, L.S., 2015. Manure, biogas digestate and crop residue management affects methane gas emissions from rice paddy fields on Vietnamese smallholder livestock farms. Nutr. Cycling Agroecosyst. 103, 329–346.
Wang, Z.Y., Xu, Y.C., Li, Z., Guo, Y.X., Wassmann, R., Neue, H.U., Lantin, R.S., Buendia, L.V., Ding, Y.P., Wang, Z.Z., 2000. A four-year record of methane emissions from irrigated rice fields in the Beijing Region of China. Nutr. Cycling Agroecosyst. 58, 55–63.
Wang, H., Dong, Y., Yang, Y., Toor, G.S., Zhang, X., 2013. Changes in heavy metal contents in animal feeds and manures in an intensive animal production region of China. J. Environ. Sci. 25, 2435–2442.
Wang, W., Lai, D.Y.F., Sardans, J., Wang, C., Datta, A., Pan, T., Zeng, C., Bartrons, M., Penuelas, J., 2015. Rice straw incorporation affects global warming potential differently in early vs. Late cropping seasons in Southeastern China. Field Crops Res. 181, 42–51.
Wassmann, R., Lantin, R.S., Neue, H.U., Buendia, L.V., Corton, T.M., Lu, Y., 2000. Characterization of methane emissions from rice fields in Asia. III. Mitigation options and future research needs. Nutr. Cycling Agroecosyst. 58, 23–36.
Watanabe, A., Takeda, T., Kimura, M., 1999. Evaluation of origins of CH4 carbon emitted from rice paddies. J. Geophys. Res. Atmos. 104, 23623–23629.
Williams, P.N., Raab, A., Feldmann, J., Meharg, A.A., 2007. Market basket survey shows elevated levels of As in South Central U.S. Processed rice compared to California: consequences for human dietary exposure. Environ. Sci. Technol. 41, 2178–2183.
Williams, P.N., Lei, M., Sun, G., Huang, Q., Lu, Y., Deacon, C., Meharg, A.A., Zhu, Y., 2009. Occurrence and Partitioning of Cadmium, Arsenic and Lead in Mine Impacted Paddy Rice: Hunan, China. Environ. Sci. Technol. 43, 637–642.
Williams, P.N., Zhang, H., Davison, W., Zhao, S., Lu, Y., Dong, F., Zhang, L., Pan, Q., 2012. Evaluation of in situ DGT measurements for predicting the concentration of Cd in chinese field-cultivated rice: impact of soil Cd:Zn ratios. Environ. Sci. Technol. 46, 8009–8016.
Xu, Y., Ge, J., Tian, S., Li, S., Nguy-Robertson, A.L., Zhan, M., Cao, C., 2015. Effects of water-saving irrigation practices and drought resistant rice variety on greenhouse gas emissions from a no-till paddy in the central lowlands of China. Sci. Total Environ. 505, 1043–1052.
Yan, X., Du, L., Shi, S., Xing, G., 2000. Nitrous oxide emission from wetland rice soil as affected by the application of controlled-availability fertilizers and mid-season aeration. Biol. Fertility Soils 32, 60–66.
Yan, X., Yagi, K., Akiyama, H., Akimoto, H., 2005. Statistical analysis of the major variables controlling methane emission from rice fields. Glob. Change Biol. Bioenergy 11, 1131–1141.
Yang, Q.W., Lan, C.Y., Wang, H.B., Zhuang, P., Shu, W.S., 2006. Cadmium in soil–rice system and health risk associated with the use of untreated mining wastewater for irrigation in Lechang. China. Agricultural Water Management 84, 147–152.
Yang, S., Peng, S., Xu, J., Luo, Y., Li, D., 2012. Methane and nitrous oxide emissions from paddy field as affected by water-saving irrigation. Phys. Chem. Earth Parts A/b/c 53, 30–37.
Yao, Z., Zheng, X., Dong, H., Wang, R., Mei, B., Zhu, J., 2012. A 3-year record of N2O and CH4 emissions from a sandy loam paddy during rice seasons as affected by different nitrogen application rates. Agric. Ecosyst. Environ. 152, 1–9.
Yu, S., Miao, Z., Xing, W., Shao, G., Jiang, Y., 2010. Research advance on irrigation and drainage for rice by using field water level as regulation index. J. Irrig. Drain. Eng. 29, 134–136.
Yuan, L., Mosley, L.M., Fitzpatrick, R., Marschner, P., 2016. Organic matter addition can prevent acidification during oxidation of sandy hypersulfidic and hyposulfidic material: effect of application form, rate and C/N ratio. Geoderma 276, 26–32.
Yuan, C., Li, F., Cao, W., Yang, Z., Hu, M., Sun, W., 2019. Cadmium solubility in paddy soil amended with organic matter, sulfate, and iron oxide in alternative watering conditions. J. Hazard. Mater. 378, 120672.
Zhang, Q., Chen, H., Huang, D., Xu, C., Zhu, H., Zhu, Q., 2019. Water managements limit heavy metal accumulation in rice: dual effects of iron-plaque formation and microbial communities. Sci. Total Environ. 687, 790–799.
Zhao, F., McGrath, S.P., Meharg, A.A., 2010. Arsenic as a food chain contaminant: mechanisms of plant uptake and metabolism and mitigation strategies. Annual Review of Plant Biology, Vol 61 (61), 535–559.
Zhou, M., Zhu, B., Brüggemann, N., Bergmann, J., Wang, Y., Butterbach-Bahl, K., 2014. N2O and CH4 emissions, and NO3− leaching on a crop-yield basis from a subtropical rain-fed wheat-maize rotation in response to different types of nitrogen fertilizer. Ecosystems 17, 286–301.
Zhu, Y., Williams, P.N., Meharg, A.A., 2008. Exposure to inorganic arsenic from rice: A global health issue? Environ. Pollut. 154, 169–171.
Zou, J., Huang, Y., Zheng, X., Wang, Y., 2007. Quantifying direct N2O emissions in paddy fields during rice growing season in mainland China: dependence on water regime. Atmos. Environ. 41, 8030–8042.
Zou, J., Huang, Y., Jiang, J., Zheng, X., Sass, R., 2005. A 3-year field measurement of methane and nitrous oxide emissions from rice paddies in China: Effects of water regime, crop residue, and fertilizer application. Global Biogeochem. Cycles 19, GB2021.
Zou, J., Liu, S., Qin, Y., Pan, G., Zhu, D., 2009. Sewage irrigation increased methane and nitrous oxide emissions from rice paddies in southeast China. Agric. Ecosyst. Environ. 129, 516–522.